\documentclass[preprint]{elsarticle}

\usepackage[utf8]{inputenc}  % UTF-8 input encoding
\usepackage[T1]{fontenc}     % Type1 fonts
\usepackage{lmodern}         % Improved Computer Modern font
\usepackage{microtype}       % Better handling of typo
\usepackage[scaled]{helvet} % Scale Helvetica
\usepackage[english]{babel}  % Hyphenation   
\usepackage{relsize} 
\usepackage{graphicx}     % For including figures
\usepackage{xcolor}       % For colored text and highlighting (optional)
\usepackage{booktabs}     % For professional-looking tables (if needed)
\usepackage{array}
\usepackage{multirow}
\usepackage[ruled,vlined,linesnumbered]{algorithm2e} % For algorithm environment
\usepackage{amsmath}      % For advanced math typesetting
\usepackage{amssymb}      % For mathematical symbols
\usepackage{mathtools}    % For additional math tools

\usepackage{caption}
\usepackage{subcaption}

\usepackage[normalem]{ulem}
\useunder{\uline}{\ul}{}
\biboptions{sort&compress}   % Sort and compress citations
\usepackage[section]{placeins}
\usepackage{tabularray}
\usepackage{longtable}
\usepackage{xcolor,hyperref}
\usepackage{tikz}

\usepackage[margin=1in]{geometry}

\begin{document}

\DeclareRobustCommand{\orcidicon}{
	\begin{tikzpicture}
	\draw[lime, fill=lime] (0,0) 
	circle [radius=0.16] 
	node[white] {{\fontfamily{qag}\selectfont \tiny ID}};
	\draw[white, fill=white] (-0.0625,0.095) 
	circle [radius=0.007];
	\end{tikzpicture}
	\hspace{-2mm}
}

\foreach \x in {A, ..., Z}{\expandafter\xdef\csname
orcid\x\endcsname{\noexpand\href{https://orcid.org/\csname orcidauthor\x\endcsname}
			{\noexpand\orcidicon}}
}

% Title

\title{Targeted Neural Architectures in Multi-Objective Frameworks for Complete Glioma Characterization from Multimodal MRI}
% Authors and affiliations
\author{Shravan Venkatraman\textsuperscript{1}, 
Pandiyaraju V\textsuperscript{1}, 
Abeshek A\textsuperscript{1}, 
Aravintakshan S A\textsuperscript{1}, 
Pavan Kumar S\textsuperscript{1}, 
Kannan A\textsuperscript{2}, 
Madhan S\textsuperscript{3}}

\author{\\ shravan.venkatraman18@gmail.com, pandiyaraju.v@vit.ac.in, abeshek.a@gmail.com, aravintcs176@gmail.com, s.pavankumar2003@gmail.com, akannan123@gmail.com, madhansingaravelu82@gmail.com}

\author{\\ \textsuperscript{1}School of Computer Science and Engineering, Vellore Institute of Technology, Chennai, India.}
\author{\\ \textsuperscript{2}Department of Information Science and Technology, College of Engineering, Guindy, Anna University, Chennai, Tamil Nadu, India.}
\author{\\ \textsuperscript{3}Department of Neurosurgery, Institute of Neurosurgery, Madras Medical College, Chennai, Tamil Nadu, India.}

\begin{frontmatter}
\begin{abstract}
Tumors in the brain are caused by abnormal growths in brain tissue resulting from different types of brain cells. If undiagnosed, they lead to aggressive neurological deficits, including cognitive impairment, motor dysfunction, and sensory loss. With the growth of the tumor, intracranial pressure will definitely increase, and this may bring about such dramatic complications as herniation of the brain, which may be fatal. Hence, early diagnosis and treatment are required to control the complications arising due to such tumors to retard their growth. Several works related to deep learning (DL) and artificial intelligence (AI) are being conducted to help doctors diagnose at an early stage by using the scans taken from Magnetic Resonance Imaging (MRI). Our research proposes targeted neural architectures within multi-objective frameworks that can localize, segment, and classify the grade of these gliomas from multimodal MRI images to solve this critical issue. Our localization framework utilizes a targeted architecture that enhances the LinkNet framework with an encoder inspired by VGG19 for better multimodal feature extraction from the tumor along with spatial and graph attention mechanisms that sharpen feature focus and inter-feature relationships. For the segmentation objective, we deployed a specialized framework using the SeResNet101 CNN model as the encoder backbone integrated into the LinkNet architecture, achieving an IoU Score of 96\%. The classification objective is addressed through a distinct framework implemented by combining the SeResNet152 feature extractor with Adaptive Boosting classifier, reaching an accuracy of 98.53\%. Our multi-objective approach with targeted neural architectures demonstrated promising results for complete glioma characterization, with the potential to advance medical AI by enabling early diagnosis and providing more accurate treatment options for patients.
\end{abstract}
\end{frontmatter}

%% ---------------------------
%% If you wish to include additional packages, define new environments or
%% new commands, put them in the file includes.tex
%%
%% Write your abstract in the file abstract.tex.
%% ---------------------------

%% ---------------------------
%% Introduction
%% ---------------------------
\textbf{Keywords}: Magnetic Resonance Imaging, Semantic Segmentation, Tumor Localization, Multimodal Image Fusion

\section{Introduction}  {
Brain tumors pose a significant challenge in both neurology and oncology, as they result from abnormal cell proliferation in the brain, affecting the central nervous system (CNS). The CNS, which consists of the brain and spinal cord, plays a crucial role in regulating responses, sensations, movements, emotions, communication, cognitive processes, and memory. The presence of brain tumors can lead to symptoms such as nausea, headaches, vomiting, and changes in hearing, vision, or speech, among other neurological impairments. These symptoms vary significantly among individuals, making early detection and appropriate treatment essential for improving patient outcomes \cite{1}.

\begin{figure}[h!]
    \centering
    \includegraphics[width=0.75\textwidth]{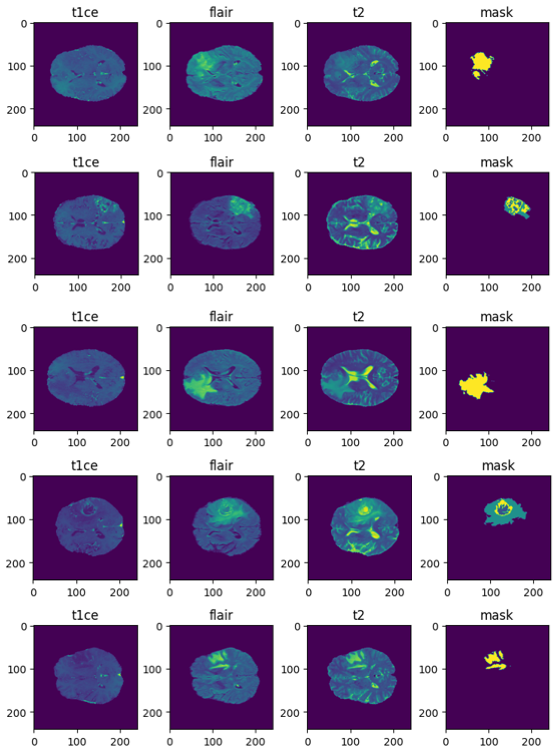}
    \caption{Segmented Ground Truth Annotations Across MRI Modalities}
    \label{fig:fig1}
\end{figure}

Tumors in the brain are broadly classified into two groups: primary and secondary. Primary tumors originate in the brain or its surrounding tissues and are further categorized into low-grade and high-grade tumors. Cells in low-grade tumors resemble normal cells and exhibit slower growth and spread compared to high-grade tumors, whereas high-grade tumors undergo rapid cell division. Secondary, or metastatic, tumors occur when cancer originating elsewhere in the body spreads to the brain. Tumors that can be distinctly segmented are sometimes easier to define, but in cases such as gliomas and glioblastomas, delineation is challenging due to their diffuse appearance, poor contrast, and extensive tentacle-like extensions \cite{2}. Figure \ref{fig:modality2} shows the heatmaps of different modalities of a 3D MRI image from multiple scales and rotations at 300 DPI (dots per inch), along with its segmentation map.

\begin{figure}[h!]
    \centering
    \includegraphics[width=0.75\textwidth]{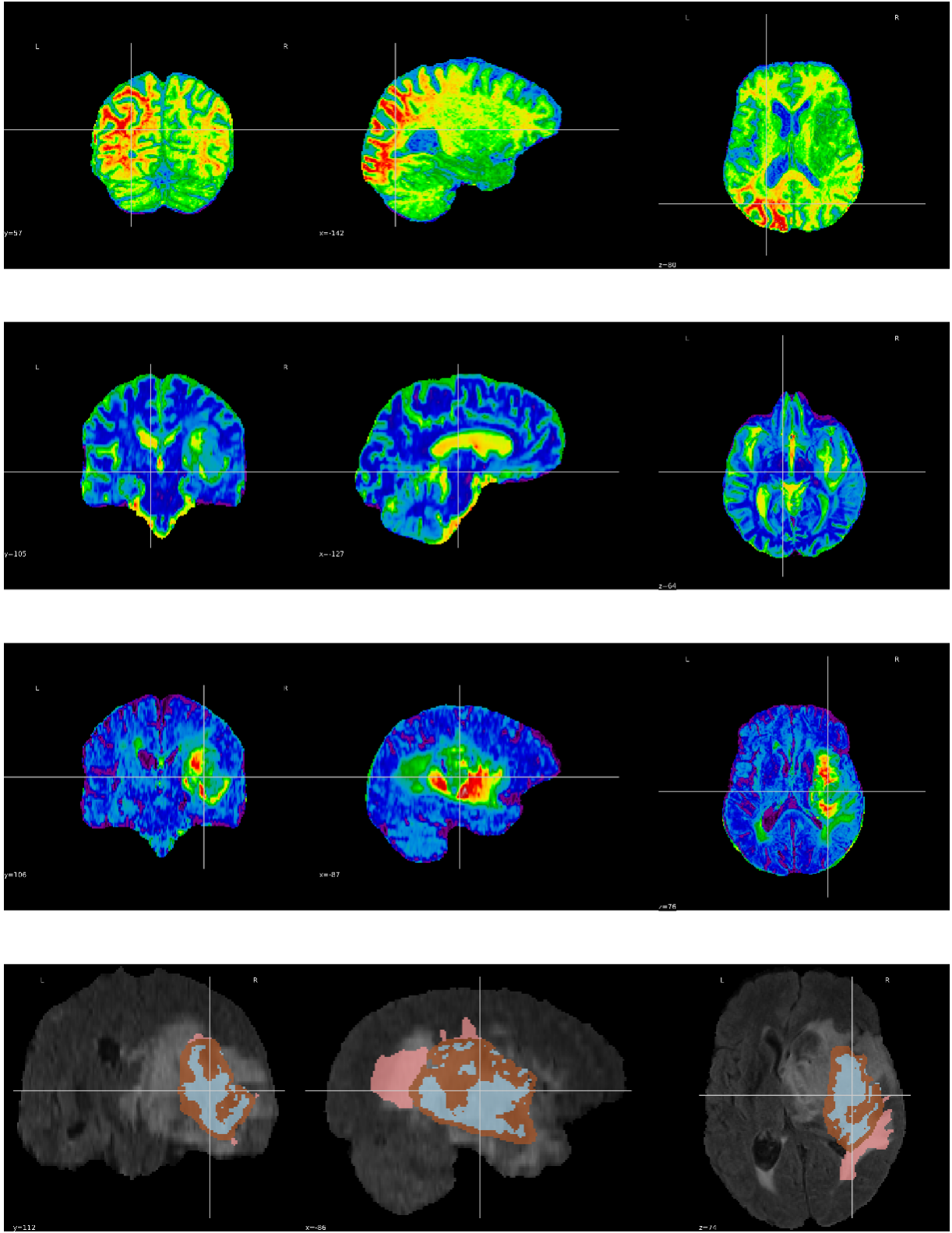}
    \caption{Multi-Modal MRI Heatmaps and Segmentation Map Visualization}
    \label{fig:modality2}
\end{figure}

Magnetic resonance imaging (MRI) is an imaging technique widely used in radiology. It produces high-resolution images of the brain and other parts of the body using strong magnetic fields and radio waves \cite{28}. Unlike other imaging modalities, MRI does not emit radiation, making it a safer option for patients. This characteristic, combined with its ability to capture fine anatomical details, makes it the preferred choice for brain imaging \cite{3}. MRI is particularly effective for accurately determining tumor location and size, as it provides superior differentiation between various brain tumor types compared to imaging techniques such as computed tomography (CT) scans. Additionally, MRI enables the visualization of soft tissues, blood vessels, and fluid accumulation, offering a comprehensive view of brain pathology.

The Brain Tumor Segmentation (BraTS) dataset was used in this research. BraTS is a widely recognized benchmark for developing and evaluating machine learning models for brain tumor detection and segmentation from MRI scans. It includes multimodal MRI sequences such as T1, T1 post-contrast (T1ce), T2, and T2 Fluid-Attenuated Inversion Recovery (FLAIR), along with expert-annotated labels for tumor regions \cite{4}. Figure~\ref{fig:fig1} presents T1ce, T2, and FLAIR MRI modalities along with their corresponding ground truth annotations, segmented separately for different scans. Figure~\ref{fig:modality2} provides heatmap-based visualizations highlighting tumor regions.

The field of deep learning (DL) has significantly advanced medical image analysis in recent years \cite{29, 30}. DL-based methods have greatly contributed to the classification and segmentation of brain tumors, offering a viable and efficient alternative to traditional approaches. Convolutional neural networks (CNNs) leverage multi-resolution image processing, mimicking the human visual system to enhance model capacity in MRI analysis. Advances in DL have led to models capable of achieving high accuracy in brain tumor classification and segmentation, making them valuable tools for medical professionals in diagnosis and treatment.

In this paper, we present a multi-objective deep learning framework for comprehensive glioma characterization from multimodal magnetic resonance imaging (MRI). Our approach integrates targeted neural architectures for precise tumor localization, high-fidelity segmentation, and glioma grading, leveraging spatial and graph attention mechanisms, deep residual networks, and adaptive ensemble classification from 3D MRI scans. For localization, we propose a spatial and graph attention-based framework capable of accurately identifying tumor subregions using multimodal features. Segmentation is performed using a hybrid LinkNet framework with a SeResNet101 backbone, generating high-precision tumor maps for further analysis. Finally, classification is achieved using a SeResNet152 backbone combined with an Adaptive Boosting classifier to distinguish between low-grade and high-grade gliomas, improving overall diagnostic accuracy.

To summarize, our contributions are as follows:
\begin{itemize}
    \item We design a hierarchical preprocessing pipeline that standardizes multimodal MRI volumes through Multiresolution Harmonic Fusion, Adaptive Focused Region Clipping, Luminance-Guided Contrast Enhancement, Dynamic Contextual Smoothing, and Statistical Feature Normalization. This pipeline enhances feature fidelity, contrast consistency, and noise artifact issues, optimizing input data for downstream tasks.
    
    \item We propose a targeted localization architecture that enhances the LinkNet framework with a VGG19-inspired encoder, integrating spatial and graph attention mechanisms to improve multimodal feature extraction and inter-feature relationships for precise glioma localization.
    
    \item We develop a specialized segmentation framework utilizing SeResNet101 as the encoder backbone within a modified LinkNet architecture, achieving an IoU score of 96\% for accurate delineation of tumor boundaries from multimodal MRI.
    
    \item We introduce a classification framework combining SeResNet152 feature extraction with Adaptive Boosting, reaching 98.53\% accuracy in glioma grading, thus completing our multi-objective approach for comprehensive tumor characterization.
\end{itemize}

The structure of this paper is as follows: Section 2 reviews the related works, and the preprocessing techniques and models of localization, segmentation, and classification are elaborated in Section 3. The experimental setup and results are then presented in Section 4, and Section 5 concludes the paper with a discussion on the findings and proposals for future directions.

 }

\section{Related Works}  {

Deep learning has significantly advanced brain tumor segmentation and classification from MRI scans. Agrawal et al. \cite{5} employed a 3D-UNet for volumetric segmentation, followed by CNN-based classification, aiming to automate diagnosis and treatment recommendation while reducing reliance on manual expertise. Their model demonstrated superior efficacy through precision and loss evaluations. Addressing segmentation challenges, Sangui et al. \cite{6} highlighted the complexity of distinguishing tumors from normal tissues due to their heterogeneous appearance. They introduced a modified U-Net architecture trained on the BRATS 2020 dataset, achieving 99.4\% accuracy and outperforming prior deep learning models in tumor segmentation.

Feng et al. \cite{7} tackled segmentation difficulties posed by gliomas' irregular shapes and textures by designing an enhanced 3D U-Net with optimized parameters, ensemble learning, and radiomic feature extraction for survival prediction. Their approach ranked 9th in the BraTS 2018 challenge and demonstrated clinical relevance by improving prognosis accuracy. Beyond segmentation, Kaur et al. \cite{8} proposed ResUNet++, integrating UNet++ with residual connections to enhance feature extraction and robustness. Similarly, Montaha et al. \cite{9} optimized U-Net for automated segmentation on 3D MRI scans, leveraging manually segmented ROIs as ground truth and exploring future enhancements through hybrid CNNs and attention mechanisms.

Saman et al. \cite{10} introduced an improved Ant-Lion Optimization algorithm to enhance feature selection from MR images, combining deep learning with handcrafted features to improve tumor classification accuracy. Wang et al. \cite{11} employed a 3D U-Net with brain-wise normalization and patching on the BraTS 2019 dataset, achieving competitive Dice scores across tumor regions. Their model further integrated survival prediction by extracting numerical features such as tumor-to-brain size ratio and surface area, marking an initial step toward clinical outcome prediction. To enhance classification accuracy, Kanchanamala et al. \cite{12} proposed an optimization-enabled hybrid DL approach that balances sensitivity and specificity in tumor detection, improving diagnostic reliability. Similarly, Ramprasad et al. \cite{13} introduced Fusion-Net and HFCMIK, leveraging probabilistic sensing for high-fidelity segmentation and classification, significantly boosting accuracy in clinical settings.

Addressing the challenge of limited annotated medical images, Isaza et al. \cite{14} applied transfer learning with data augmentation to improve tumor detection in MRI scans. Kim et al. \cite{15} tackled annotation constraints through an active learning framework that employs uncertainty sampling, redundancy reduction, and optimized data initialization, enhancing segmentation efficiency with minimal labeled data. Nodirov et al. \cite{16} advanced 3D medical image segmentation by integrating skip connections and attention modules within a 3D U-Net model. Their architecture incorporates pre-trained 3D MobileNetV2 blocks for computational efficiency and rapid convergence while leveraging attention mechanisms to filter irrelevant features, optimizing tumor segmentation performance.

Dang et al. \cite{17} introduced a framework integrating image preprocessing techniques with a DL model to enhance segmentation and classification accuracy. Their approach improved feature extraction, leading to better tumor identification. Shaukat et al. \cite{18} addressed glioma segmentation challenges by proposing a cloud-based 3D U-Net framework, enabling remote training and accessibility across various terminal devices. Their model achieved a 95\% accuracy, emphasizing the role of DL in making tumor detection more scalable and accessible.

Kollem et al. \cite{19} developed a novel DL architecture for improved tumor detection and classification, leveraging advanced feature extraction techniques to enhance reliability and precision. Aswani et al. \cite{20} further refined feature optimization by introducing a dual autoencoder with Singular Value Decomposition (SVD). This hybrid approach extracted and refined significant features, improving segmentation accuracy and aiding in precise diagnosis and treatment planning.

\subsection{Research Gap}

Despite significant advancements in deep learning for brain tumor segmentation and classification, several challenges remain unaddressed. Existing segmentation models, such as 3D U-Net and its variants \cite{6,7,18}, have achieved high accuracy but often struggle with precise glioma localization due to insufficient multimodal feature integration and limited spatial awareness. These models lack dedicated attention mechanisms to enhance feature relationships across modalities, leading to suboptimal localization accuracy.

Furthermore, while frameworks such as ResUNet++ \cite{8} and modified U-Net architectures \cite{9,19} have improved segmentation precision, they primarily focus on conventional encoder-decoder structures without leveraging advanced backbone architectures. The absence of targeted feature extraction modules, such as deep residual networks, limits the ability to capture fine-grained tumor boundaries effectively.

In classification, existing methods \cite{10,20} predominantly rely on conventional CNNs or handcrafted feature selection, often failing to leverage ensemble-based approaches for enhanced robustness. While methods like Ant-Lion Optimization \cite{10} have improved feature selection, they do not integrate strong deep feature representations from high-capacity architectures, restricting their generalization to diverse glioma subtypes.

To address these gaps, we propose a targeted localization architecture integrating spatial and graph attention mechanisms within an enhanced LinkNet framework, leveraging a VGG19-inspired encoder for multimodal feature extraction. Additionally, we introduce a segmentation framework incorporating SeResNet101 as the encoder backbone, achieving high-precision tumor boundary delineation with an IoU score of 96\%. Finally, our classification model integrates SeResNet152 with Adaptive Boosting, significantly improving glioma grading accuracy (98.53\%) by combining deep feature extraction with ensemble learning.

}
\section{Methodology} \label{sec:DS}  {
\subsection{\textbf{Overall System Workflow}}
% fig 
\begin{figure}[h!]
    \centering
    \includegraphics[width=0.9\textwidth]{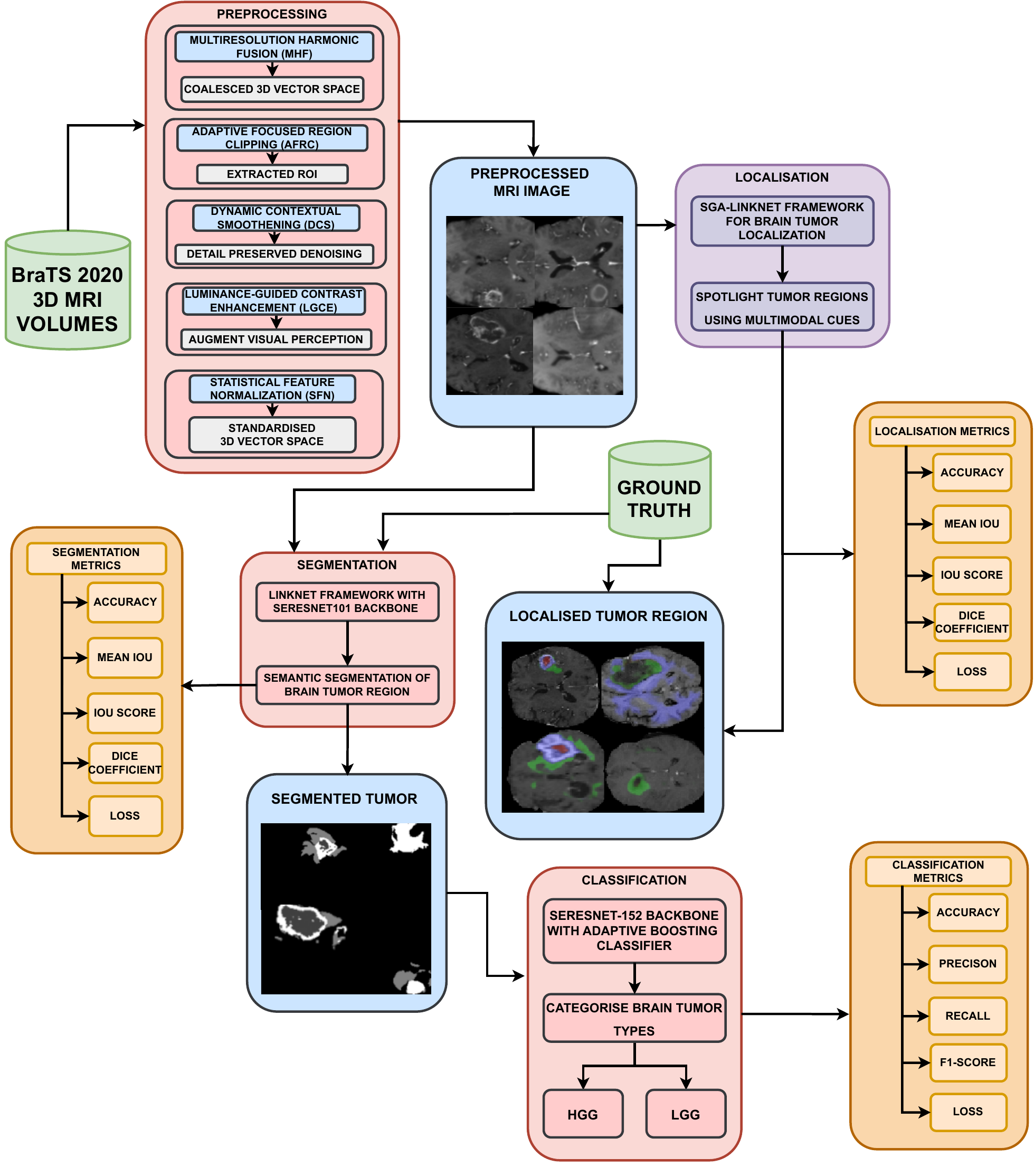}
    \caption{Overall Workflow of the Proposed System Architecture}
    \label{fig:overallArchitecture}
\end{figure}
Initially, 3D MRI images from the BraTS 2020 dataset undergo a preprocessing pipeline comprising of a series of image processing methods. Following preprocessing, the images are individually input into two distinct frameworks for localization and segmentation. The SGA-LinkNet Localization Framework identifies tumor regions in MRI scans using multimodal cues that delineate subregions such as "edema/invasion", "non-enhancing", "enhancing", and "necrosis". Simultaneously, the LinkNet Segmentation Framework with a SeResNet101 CNN backbone performs semantic segmentation, from which detailed segmentation maps of the tumor region from the MRI images are obtained. This is passed through the SeResNet-152 CNN Architecture with an Adaptive Boosting classifier, classifying tumors into either of the two categories: high-grade glioma (HGG) or low-grade glioma (LGG). All localization and segmentation effectiveness tests are conducted using the aforementioned metrics: accuracy, average intersection over union (IoU) score, mean IoU score, Dice coefficient, and a global loss function that combines focal loss and Dice loss. Classification effectiveness is evaluated using metrics such as F1-score, accuracy, recall, precision, and binary cross-entropy loss. Figure~\ref{fig:overallArchitecture} illustrates the comprehensive workflow of the proposed system architecture.

\subsection{\textbf{Dataset Description}}  
We utilize the Multimodal Brain Tumor Segmentation Challenge 2020 (BraTS2020) dataset [\cite{21}, \cite{22}, \cite{23}, \cite{24}, \cite{25}] in this research. This dataset consists of 3D MRI scans of glioblastoma (GBM)/high-grade glioma (HGG) and low-grade glioma (LGG) across multiple modalities.  

The dataset categorizes tumors into four distinct classes:  
\begin{itemize}
    \item \textbf{Necrosis}: Represents the necrotic core of the tumor, consisting of dead tissue.  
    \item \textbf{Edema}: Refers to the tumor-induced swelling caused by fluid accumulation.  
    \item \textbf{Enhancing Tumor}: Comprises tumor regions with increased vascularity, which fluoresce more prominently in contrast-enhanced images.  
    \item \textbf{Non-Enhancing Tumor}: Represents tumor regions that do not exhibit contrast enhancement.  
\end{itemize}  

We utilize native T1 (T1), T2-weighted (T2), and T2 Fluid-Attenuated Inversion Recovery (FLAIR) MRI scans as they provide complementary structural and pathological information necessary for precise tumor characterization. T1 scans offer detailed anatomical contrast, while T2 and FLAIR sequences enhance the visibility of peritumoral regions, such as edema and infiltrative tumor components. These modalities collectively improve segmentation accuracy and facilitate comprehensive glioma analysis. Table~\ref{tab:brats_statistics} provides the distribution of LGG and HGG cases across MRI modalities. The dataset is split into 80\% for training and 20\% for validation, ensuring robust model generalization and performance evaluation on unseen data.

\begin{table}[h!]
\centering
\caption{Distribution of Tumor Types in Training and Validation Data.}
\label{tab:brats_statistics}
\begin{tabular}{lcccccc}
\toprule
\textbf{Tumor Type} & \multicolumn{3}{c}{\textbf{Training Data}} & \multicolumn{3}{c}{\textbf{Validation Data}} \\ 
\cmidrule(lr){2-4} \cmidrule(lr){5-7} 
& \textbf{T1} & \textbf{T2} & \textbf{Flair} & \textbf{T1} & \textbf{T2} & \textbf{Flair}  \\ 
\hline
High Grade Glioma (HGG) & 233 & 233 & 233 & 58 & 58 & 58 \\
Low Grade Glioma (LGG)  & 62  & 62  & 62  & 16 & 16 & 16 \\
\midrule
\textbf{Total}          & 295 & 295 & 295 & 74 & 74 & 74 \\
\bottomrule
\end{tabular}
\end{table}

\subsection{Preprocessing}

\begin{algorithm}[!h]
\caption{Hierarchical MRI Preprocessing Pipeline}
\label{algo:mri_preprocessing}
\KwIn{MRI sequences: T1CE, FLAIR, T2 volumes $\{X_{\text{T1CE}}, X_{\text{FLAIR}}, X_{\text{T2}}\} \in \mathbb{R}^{H\times W\times D}$; wavelet basis $\mu$; gamma correction factor $\gamma$; median-filter kernel size $k$; cropping parameters $s_{\text{ROI}}, s_d$; margin $r$}
\KwOut{Preprocessed MRI volume $X_{\text{pre}}\in [0,1]^{H'\times W'\times D'}$}

\textbf{Step 1: Multiresolution Harmonic Fusion (MHF)}
Convert 3D volumes to 2D slices:  
\[
X_{m}^{(2D)} \gets \text{slice\_conversion}(X_m), \quad m\in\{\text{T1CE, FLAIR, T2}\}
\]

Apply Discrete Wavelet Transform (DWT) at scale $j$:  
\[
(a_m^j, d_m^j) \gets \text{DWT}(X_{m}^{(2D)}, \mu), \quad m\in\{\text{T1CE, FLAIR, T2}\}
\]

Fuse approximation and detail coefficients:  
\[
a_f \gets \frac{1}{3} \sum_{m} a_m^j, \quad d_f \gets \frac{1}{3} \sum_{m} d_m^j
\]

Reconstruct fused image using inverse DWT:  
\[
X_f \gets \text{iDWT}(a_f, d_f, \mu)
\]

\textbf{Step 2: Adaptive Focused Region Clipping (AFRC)}
Extract ROI using spatial cropping:  
\[
X_f \gets X_f[s_{\text{ROI}}:H-s_{\text{ROI}},\, s_{\text{ROI}}:W-s_{\text{ROI}},\, s_d:D-s_d]
\]

\textbf{Step 3: Luminance-Guided Contrast Enhancement (LGCE)}
Apply voxel-wise gamma correction:  
\[
X_f(i,j,k) \gets 255 \cdot \left( \frac{X_f(i,j,k)}{255} \right)^\gamma
\]

\textbf{Step 4: Statistical Feature Normalization (SFN)}
Normalize intensities to $[0,1]$:  
\[
X_f \gets \frac{X_f - \min(X_f)}{\max(X_f) - \min(X_f)}
\]

\textbf{Step 5: Dynamic Contextual Smoothing (DCS)}
Apply median filter $\mathcal{M}_k$:  
\[
X_f \gets \mathcal{M}_k(X_f)
\]

\textbf{Step 6: Final Resizing and Normalization}
Remove margin $r$:  
\[
X_{\text{pre}} \gets X_f[r:H-r,\,r:W-r,\,r:D-r]
\]

Normalize final volume:  
\[
X_{\text{pre}} \gets \frac{X_{\text{pre}} - \min(X_{\text{pre}})}{\max(X_{\text{pre}}) - \min(X_{\text{pre}})}
\]

\Return $X_{\text{pre}}$
\end{algorithm}

\subsubsection{Multiresolution Harmonic Fusion (MHF)}

The proposed Multiresolution Harmonic Fusion (MHF) technique constructs a harmonized 3D vector space by leveraging multiscale frequency decomposition and selective reconstruction. This approach decomposes a 3D MRI volume into multiple 2D sub-bands, processes them independently, and fuses them back into a high-resolution 3D representation, enhancing feature distinctiveness and structural clarity.

Given a 3D image \( I(x,y,z) \), MHF applies a multiresolution wavelet decomposition:

\begin{equation}
I(a,b,c) \xrightarrow{\text{MHF Decomposition}} \{ I_{LL}(a,b), I_{LH}(a,b), I_{HL}(a,b), I_{HH}(a,b) \}_z
\end{equation}

\[
\underbrace{\text{LL}}_{\text{low-low frequency}}, \ 
\underbrace{\text{LH}}_{\text{low-high frequency}}, \ 
\underbrace{\text{HL}}_{\text{high-low frequency}}, \ 
\underbrace{\text{HH}}_{\text{high-high frequency}}
\]

For optimal frequency-domain analysis, we employ a biorthogonal 1.3 (bior 1.3) wavelet, wherein decomposition utilizes a length-1 analysis filter, and reconstruction employs a length-3 synthesis filter. This results in an eight-band vector representation, where high-frequency components capture fine-grained details, while low-frequency coefficients preserve structural consistency.

To achieve effective fusion, harmonic averaging is applied to wavelet coefficients across decomposed sub-bands:

\begin{equation}
C_{\text{harmonic}}(a,b,c) = 
\underbrace{\frac{1}{N}}_{\text{Normalization Factor}} 
\sum_{i=1}^N 
\underbrace{C_i(a,b,c)}_{\text{coefficients from the } i\text{-th decomposed 2D sub-band}}
\end{equation}

Finally, the fused coefficients undergo inverse wavelet transformation to reconstruct the enhanced MRI volume:

\begin{equation}
I_{\text{harmonic}}(a,b,c) \xrightarrow{\text{Inverse MHF Transform}} I_{\text{enhanced}}(a,b,c)
\end{equation}

\subsubsection{Adaptive Focused Region Clipping (AFRC)}

To enhance focus on the tumor region, Adaptive Focused Region Clipping (AFRC) removes non-contributory areas from the 3D MRI tensor. A uniform resizing factor is applied across all dimensions to retain only the Region of Interest (ROI), reducing irrelevant information and improving model performance.

Let \( I_{\text{harmonic}}(x,y,z) \) be the preprocessed 3D MRI tensor. Using a resizing factor \( \alpha \), AFRC extracts the ROI by:

\begin{equation}
I'(x,y,z) = 
\begin{cases} 
I_{\text{harmonic}}(x,y,z) & \text{if } |x - x_{\text{center}}| \leq \alpha_{\text{size}_x}, |y - y_{\text{center}}| \leq \alpha_{\text{size}_y}, |z - z_{\text{center}}| \leq \alpha_{\text{size}_z} \\
0 & \text{otherwise}
\end{cases}
\end{equation}

This ensures only the critical region is retained for segmentation and classification.

\subsubsection{Dynamic Contextual Smoothing (DCS)}

Dynamic Contextual Smoothing (DCS) reduces salt-and-pepper noise in MRI scans using a non-linear median-based approach. A 3D median kernel (\(3 \times 3 \times 3\)) scans through the image, replacing each voxel with the median of its local neighborhood, preserving edge structures while eliminating noise.

\begin{equation}
I_{\text{smoothed}}(a,b,c) = \underbrace{\text{median}}_{\text{Median Operator}} \left( \underbrace{I_{\text{input}}(a+p,b+q,c+r) \mid p,q,r \in W}_{\text{Neighborhood Voxels}} \right)
\end{equation}

\subsubsection{Luminance-Guided Contrast Enhancement (LGCE)}

Luminance-Guided Contrast Enhancement (LGCE) adjusts brightness and contrast by applying a power-law transformation to voxel intensities. A gamma correction factor \( \gamma \) modifies the filtered image, enhancing structural details while preserving intensity relationships.

\begin{equation}
I_{\text{enhanced}}(p,q,r) = \underbrace{A}_{\text{Normalization Factor}} \cdot \underbrace{I_{\text{smoothed}}(p,q,r)^\gamma}_{\text{Gamma-Corrected Intensity}}
\end{equation}

\subsubsection{Statistical Feature Normalization (SFN)}

Statistical Feature Normalization (SFN) standardizes voxel intensity values to the range [0,1], stabilizing model training and improving convergence. This is achieved by scaling intensities relative to the maximum possible voxel value.

\begin{equation}
\hat{I}(p,q,r) = \frac{I_{\text{enhanced}}(p,q,r)}{255}
\end{equation}

This preprocessing pipeline ensures noise reduction, enhanced contrast, and intensity normalization, optimizing MRI scans for accurate segmentation and classification. Figure~\ref{fig:preprocessingSteps} depicts the outputs obtained after each step in our MRI image preprocessing pipeline. The algorithmic workflow for our MRI image preprocessing pipeline is detailed in Algorithm~\ref{algo:mri_preprocessing}.

% fig
\begin{figure}[h!]
    \centering
    \includegraphics[width=0.8\textwidth]{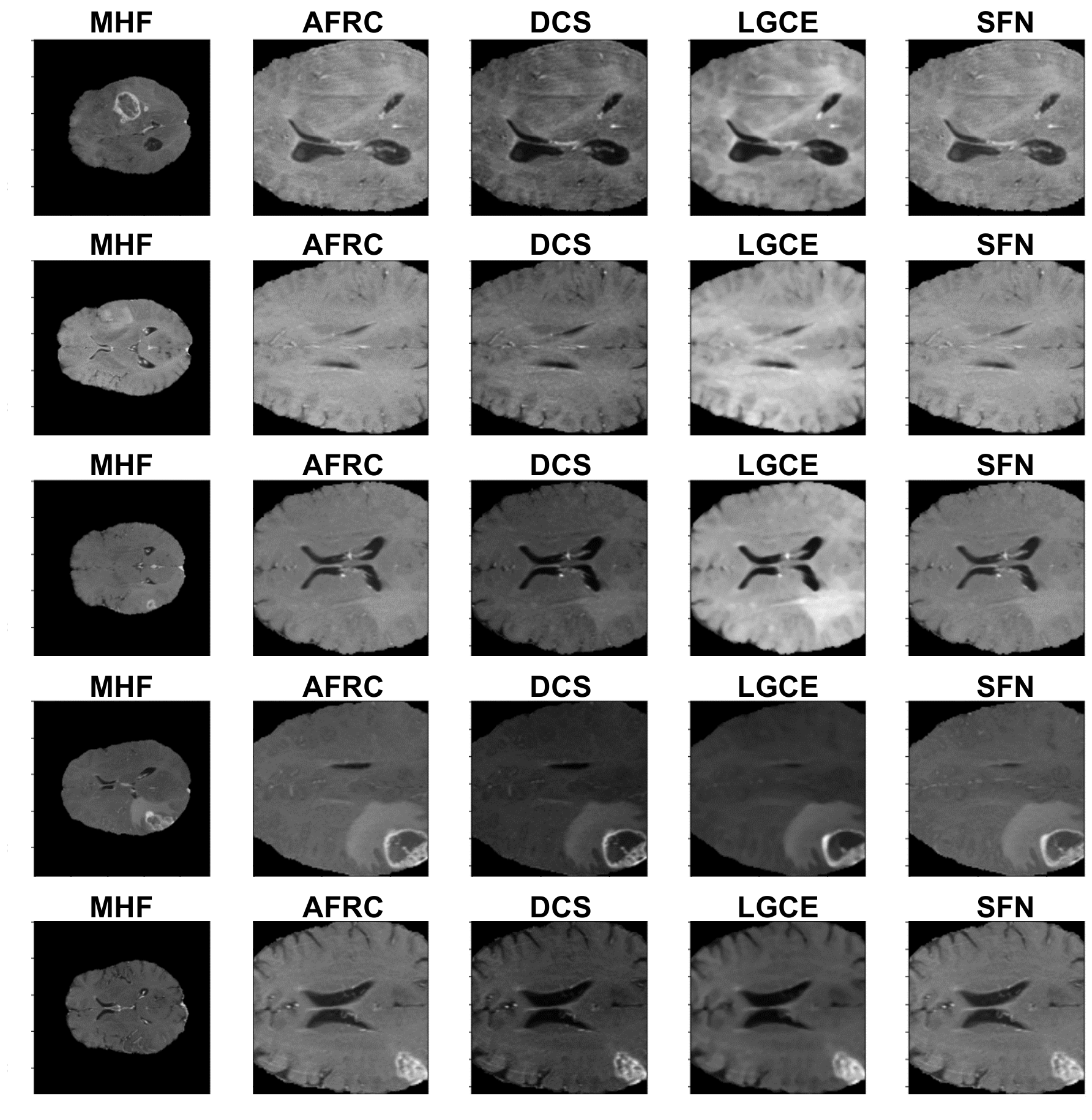}
    \caption{MRI volume outputs after each step in our preprocessing pipeline.}
    \label{fig:preprocessingSteps}
\end{figure}

% fig
\begin{figure}[h!]
    \centering
    \includegraphics[width=\textwidth]{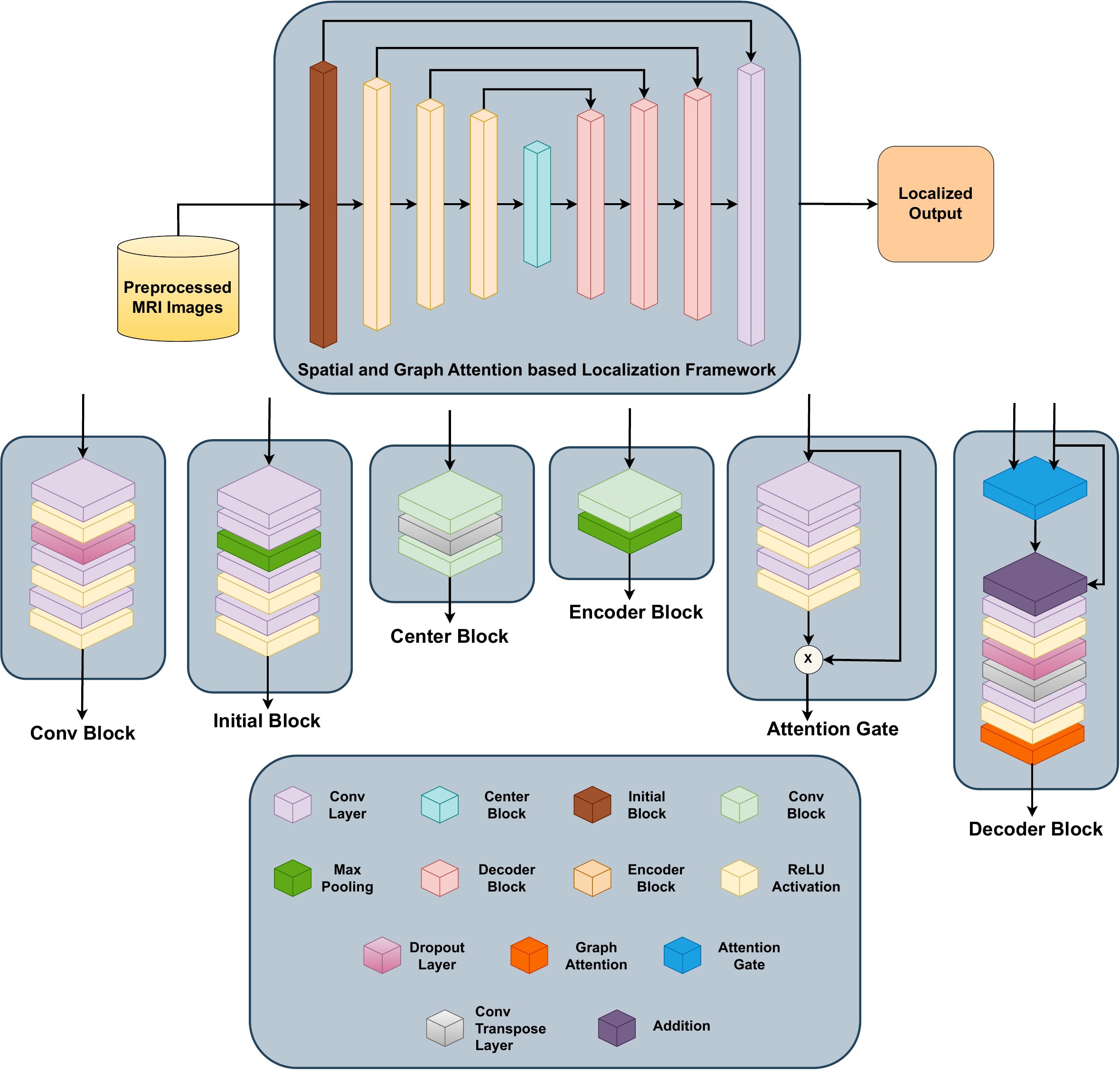}
    \caption{Spatial and Graph Attention based LinkNet Framework (SGA LinkNet) for Brain Tumor Localization}
    \label{fig:localizationArchitecture}
\end{figure}

\subsection{\textbf{Spatial-Graph Attention Enhanced LinkNet Framework for Brain Tumor Localization}}

\begin{algorithm}[t]
\caption{SGA-LinkNet for Brain Tumor Localization}
\label{algo:sga_linknet_localization}
\KwIn{MRI volume $X \in \mathbb{R}^{H\times W\times D}$, model parameters $\theta$}
\KwOut{Tumor localization map $\hat{Y}\in[0,1]^{H\times W\times D}$}

\tcp{Encoder Stage}
$X^{(0)} \gets X$\;
\For{$l=1,\dots,L$ encoder layers}{
    $X^{(l)} \gets \text{ReLU}(\text{Conv}(X^{(l-1)}; W_e^{(l)}, b_e^{(l)}))$\;
    $X^{(l)} \gets \text{MaxPool}(X^{(l)})$\;
}

\tcp{Center Block (Global Context Extraction)}
$X^c \gets \text{ConvBlock}(X^{(L)})$\;
$X^c \gets \text{Conv}(\text{ReLU}(X^c))$\;
$X^c \gets \text{ConvBlock}(X^c)$\;

\tcp{Decoder Stage with Attention-based Skip Connections}
\For{$l=L,\dots,1$ decoder layers}{
    \tcp{Spatial Attention Fusion}
    $A^{(l)} \gets \sigma\left(f_s(X^{(l)}) + f_p(X^c)\right)$\;
    $X^d \gets A^{(l)} \odot X^{(l)} + X^c$\;

    \tcp{Upsampling and Refinement}
    $X^d \gets \text{TransposeConv}(X^d)$\;
    $X^d \gets \text{ConvBlock}(X^d)$\;

    \tcp{Graph Attention Refinement}
    \For{each node feature $h_i\in X^d$}{
        Compute attention weights: 
        $\alpha_{ij} = \frac{\exp\left(\text{LeakyReLU}\left(a^T[Wh_i\|Wh_j]\right)\right)}{\sum_{k\in\mathcal{N}_i}\exp\left(\text{LeakyReLU}(a^T[Wh_i\|Wh_k])\right)}$\;
        Aggregate neighborhood features:
        $h_i' \gets \sum_{j\in \mathcal{N}_i}\alpha_{ij}(Wh_j)$\;
    }
    $X^c \gets X^d$\;
}

\tcp{Final Localization Prediction}
$\hat{Y} \gets \text{Softmax}(\text{Conv}(X^d))$\;

\Return $\hat{Y}$\;
\end{algorithm}

We introduce a Spatial and Graph Attention enhanced LinkNet (SGA-LinkNet) for precise localization of brain tumor regions using multimodal MRI data. SGA-LinkNet leverages both hierarchical convolutional features and attention-driven refinement to highlight tumor regions with high spatial accuracy. The proposed architecture extends the LinkNet~\cite{26} model by integrating spatial attention mechanisms and Graph Attention Networks (GATs) directly into the decoder stages, significantly improving localization performance.

The encoder of our model captures hierarchical spatial features through successive convolutional blocks inspired by the VGG19 architecture~\cite{27}. Initially, the input MRI volumes $X \in \mathbb{R}^{H \times W \times D}$ pass through an initial convolutional block composed of two consecutive convolutional layers, each followed by a ReLU activation and a max-pooling layer, effectively reducing the spatial dimension while enriching feature representations. Subsequent encoder blocks further process these features through sequences of convolutional layers, each performing convolution operations of the form $X' = \text{ReLU}(X \ast W + b)$, where $W$ and $b$ denote learnable weights and biases. Dropout layers are strategically interspersed to mitigate overfitting, resulting in the hierarchical representations essential for accurate localization.

The central encoder block integrates two additional convolutional stages, enhancing global context modeling. Given the input $X$, it performs sequential convolutions, producing intermediate representations $X'' = \text{ConvBlock}(X)$, followed by a refinement step yielding $X''' = \text{ConvBlock}(\text{Conv}(X''))$. This design enriches spatial context by capturing complex interdependencies among deeper convolutional features.

In the decoder, we employ spatial attention mechanisms at each decoding stage to selectively fuse skip-connected encoder features with decoder features. Formally, given spatial encoder features $X_s$ and positional decoder features $X_p$, the spatial attention map is computed as $A = \sigma(f_s(X_s) + f_p(X_p))$, where $\sigma$ denotes the sigmoid activation, and $f_s, f_p$ are convolutional operations. The attention-refined feature maps $X_d$ are subsequently upsampled using transpose convolutions defined as $X_u(i,j,k) = \sum_{p,q,c} X_d(i+s\cdot p, j+s\cdot q, c) \cdot K(p,q,c,d)$, where $K$ is the learned kernel and $s$ is the stride, effectively restoring spatial resolution.

We then refine these spatially attended features performed through Graph Attention Networks, enabling the modeling of semantic relationships among localized regions. Specifically, given feature nodes $h_i$ and $h_j$, the graph attention mechanism computes attention coefficients $\alpha_{ij}$ as:

\begin{equation}
\alpha_{ij} = \frac{\exp\left(\text{LeakyReLU}\left(a^T [W h_i \parallel W h_j]\right)\right)}{\sum_{k \in \mathcal{N}_i}\exp\left(\text{LeakyReLU}\left(a^T [W h_i \parallel W h_k]\right)\right)},
\end{equation}
where $W$ and $a$ are learned weights, and $\parallel$ denotes concatenation. The attention-weighted node representations are then aggregated as $h'_i = \sum_{j \in \mathcal{N}_i}\alpha_{ij}(W h_j)$, effectively capturing contextual relationships and reinforcing tumor-specific regions. This ensures that spatially and semantically similar labels are treated cohesively by applying smoothness constraints because the segmentation labels assigned to the various tumor regions in the MRI scans can be considered as nodes, while the relationships between these nodes can be regarded as edges. Consistent localization maps are guided by the model for producing such ones that respect anatomical and functional boundaries through the nodes representing adjacent or functionally related brain regions that might be connected.

Finally, the decoder outputs localization predictions via a convolutional layer activated by a softmax function, defined as :
\begin{equation}
\hat{y}(p) = \frac{\exp(p)}{\sum_{(i,j,k)} \exp(p_{ijk})}
\end{equation}
which generates probabilistic tumor maps highlighting distinct localized tumor regions. The proposed SGA-LinkNet localization framework thus effectively integrates spatial and graph-based attention mechanisms, leveraging hierarchical convolutional features for precise tumor localization in multimodal brain MRI. The model specifications and parameters of the proposed SGA-LinkNet Framework for Brain Tumor Localization are shown in Table~\ref{tab:SGA_Specs}. Algorithm~\ref{algo:sga_linknet_localization} outlines the working of our proposed localization framework. 

% algo 2

\begin{table}[h!]
\centering
\caption{SGA LinkNet Brain Tumor Localization Framework Specifications}
\begin{tabular}{lcc}
\hline
\textbf{Parameters} & \textbf{Coefficients} \\ \hline
Epochs              & 100                  \\
Batch Size          & 8                    \\ 
Learning Rate       & 0.001                \\ 
Total Trainable Parameters & 64,436,912        \\ 
Image Shape         & (128, 128, 128)      \\ \hline
\end{tabular}
\label{tab:SGA_Specs}
\end{table}

\subsection{\textbf{SE-ResNet101-based LinkNet for Tumor Segmentation}}
We propose a 3D brain tumor segmentation framework leveraging a LinkNet \cite{26} architecture enhanced with SE-Residual blocks derived from SeResNet101 as the encoder backbone. The core innovation is integrating Squeeze-and-Excitation (SE) mechanisms directly into residual blocks, forming SE-Residual blocks. This approach effectively recalibrates channel-wise features, significantly enhancing segmentation accuracy. This recalibration process is particularly important for brain tumor segmentation, as it mimics the radiologists' attentional focus on specific tumor-relevant features while suppressing non-informative background tissue variations.

Specifically, given a 3D MRI input volume $X \in \mathbb{R}^{H \times W \times D \times C}$, the encoder progressively extracts features using SE-Residual blocks. Each SE-Residual block performs residual learning with channel-wise attention recalibration, expressed as 
\begin{equation}
Y_{\text{enc}} = [F(X, \mathbf{W}_i) + X] \odot \sigma(S(X)),
\end{equation}
where $F(X, \mathbf{W}_i)$ is the residual function, and $S(X)$ denotes the SE block output. This mechanism parallels the biological visual processing pathway where certain features (analogous to specific tumor textures, intensity patterns, and boundaries) are selectively emphasized over others. The residual connections preserve critical information about normal brain anatomy, which serves as a reference for identifying pathological deviations.

The hierarchical feature extraction in our encoder mirrors the hierarchical nature of brain tumor composition, from microscopic cellular abnormalities to macroscopic tissue changes. Lower-level features capture local intensity variations (corresponding to cellular density changes), while higher-level features represent complex structural patterns (reflecting tumor morphology and infiltration patterns). The SE mechanism dynamically weights these features according to their relevance for differentiating tumor from healthy tissue.

The decoder operates by reconstructing spatial resolutions from deep encoder features. At each decoder stage $l$, feature maps $Z^{(l)}$ are processed via convolutional operations and upsampled: 

\begin{equation}
Y_{\text{dec}}^{(l)} = \phi(U(\mathbf{W}_d^{(l)} \ast Z^{(l)} + \mathbf{b}_d^{(l)}))
\end{equation}, 

where $\mathbf{W}_d^{(l)}$ and $\mathbf{b}_d^{(l)}$ represent learned weights and biases. This reconstruction process is analogous to the integration of diverse radiological findings that diagnosticians perform when delineating tumor boundaries.

Each decoder stage output $Y_{\text{dec}}^{(l)}$ is combined with its corresponding encoder output $Y_{\text{enc}}^{(l)}$ via an additive skip connection: 

\begin{equation}
Y_{\text{skip}}^{(l)} = Y_{\text{enc}}^{(l)} + Y_{\text{dec}}^{(l)}
\end{equation}

These skip connections are crucial for preserving fine anatomical details that might otherwise be lost during downsampling, much like how radiologists simultaneously consider both local tissue characteristics and broader contextual information when identifying tumor regions. This is particularly important for accurate segmentation of tumor boundaries where infiltrative growth patterns create subtle transitions between healthy and pathological tissue.

Our architecture's element-wise addition approach for feature fusion is especially suitable for gliomas and other infiltrative brain tumors, where the intermingling of tumor cells with normal brain parenchyma creates complex transition zones. Unlike concatenation-based approaches, addition-based fusion better represents the overlapping nature of these transition zones by integrating features rather than merely juxtaposing them.

Finally, the aggregated skip-connected feature map is passed through a convolutional output layer to yield the final segmentation prediction $\hat{Y} = \psi(\mathbf{W}_{\text{out}} \ast Y_{\text{skip}}^{(L)} + \mathbf{b}_{\text{out}})$, mapping learned feature representations into tumor segmentation masks. The resulting SE-ResNet101-based LinkNet framework effectively segments tumor regions by effectively modeling both the distinctive appearances of different tumor components across multiple MRI sequences and the spatial relationships between these components, which typically follow known patterns of glioma growth and invasion along white matter tracts and other preferential pathways. Our segmentation framework's specifications and parameters for brain tumor segmentation are listed in Table~\ref{tab:segSpecs}.

\begin{algorithm}[!t]
\caption{Training SE-ResNet101-based LinkNet for Brain Tumor Segmentation}
\label{algo:seresnet101_linknet_segmentation}
\KwIn{Preprocessed MRI volume $X\in\mathbb{R}^{H\times W\times D}$, ground-truth segmentation mask $Y$, initial learning rate $\mu$, focal loss parameters $\alpha,\gamma$, patience $\tau$, LR reduction factor $f$, epochs $E$}
\KwOut{Optimized SE-ResNet101-LinkNet parameters $\theta$}

Initialize parameters $\theta$ randomly; best validation loss $L_{\text{best}}\gets\infty$, stagnation counter $c\gets 0$\;

\For{$epoch = 1,\dots,E$}{
    Sample a training batch $(X_b,Y_b)$\;

    \tcp{Forward Pass: Encoder (SE-Residual Blocks)}
    $X_{\text{enc}}^{(0)}\gets X_b$\;
    \For{$l=1,\dots,L$ encoder blocks}{
        $X_{\text{enc}}^{(l)}\gets [F(X_{\text{enc}}^{(l-1)};W_l) + X_{\text{enc}}^{(l-1)}]\cdot S(X_{\text{enc}}^{(l-1)})$\;
    }

    \tcp{Center Block (Contextual Refinement)}
    $X_c\gets\text{ConvBlock}(X_{\text{enc}}^{(L)})$\;

    \tcp{Decoder Stage with Skip Connections (Feature Fusion)}
    $X_{\text{dec}}\gets X_c$\;
    \For{$l=L,\dots,1$ decoder blocks}{
        $X_{\text{dec}}\gets\text{Upsample}(\text{Conv}(X_{\text{dec}}))$\;
        $X_{\text{dec}}\gets\text{ReLU}(X_{\text{dec}})$\;
        $X_{\text{dec}}\gets X_{\text{dec}} + X_{\text{enc}}^{(l)}$\;
    }

    \tcp{Segmentation Prediction}
    $\hat{Y}_b\gets\text{Softmax}(\text{Conv}(X_{\text{dec}}))$\;

    \tcp{Loss Computation}
    Compute focal loss:
    $L_{\text{focal}}\gets -\alpha(1-\hat{Y}_b)^\gamma Y_b\log(\hat{Y}_b)-\alpha\hat{Y}_b^\gamma(1-Y_b)\log(1-\hat{Y}_b)$\;

    Compute Jaccard loss (IoU-based):
    $L_{\text{jaccard}}\gets 1 - \frac{|\hat{Y}_b\cap Y_b|}{|\hat{Y}_b\cup Y_b|}$\;

    Total loss:
    $L_{\text{total}}\gets L_{\text{focal}} + L_{\text{jaccard}}$\;

    \tcp{Parameter Update}
    Update parameters via gradient descent:
    $\theta\gets\theta-\mu\nabla_{\theta}L_{\text{total}}$\;

    Evaluate validation loss $L_{\text{val}}$\;
    \uIf{$L_{\text{val}} < L_{\text{best}}$}{
        $L_{\text{best}}\gets L_{\text{val}}$, $c\gets 0$\;
    }
    \Else{
        $c\gets c+1$\;
        \If{$c\geq\tau$}{
            $\mu\gets\mu/f$, $c\gets0$\;
        }
    }
}

\Return optimized model parameters $\theta$\;
\end{algorithm}

\begin{table}[h!]
\centering
\caption{LinkNet Segmentation Framework with SeResNet101 Backbone Specifications}
\begin{tabular}{lcc}
\hline
\textbf{Parameters} & \textbf{Coefficients} \\ \hline
Epochs              & 100                  \\
Batch Size          & 8                    \\ 
Learning Rate       & 0.0001               \\ 
Total Trainable Parameters & 94,673,582        \\ 
Image Shape         & (128, 128, 128)      \\ \hline
\end{tabular}
\label{tab:segSpecs}
\end{table}

% fig
\begin{figure}[h!]
    \centering
    \includegraphics[width=\textwidth]{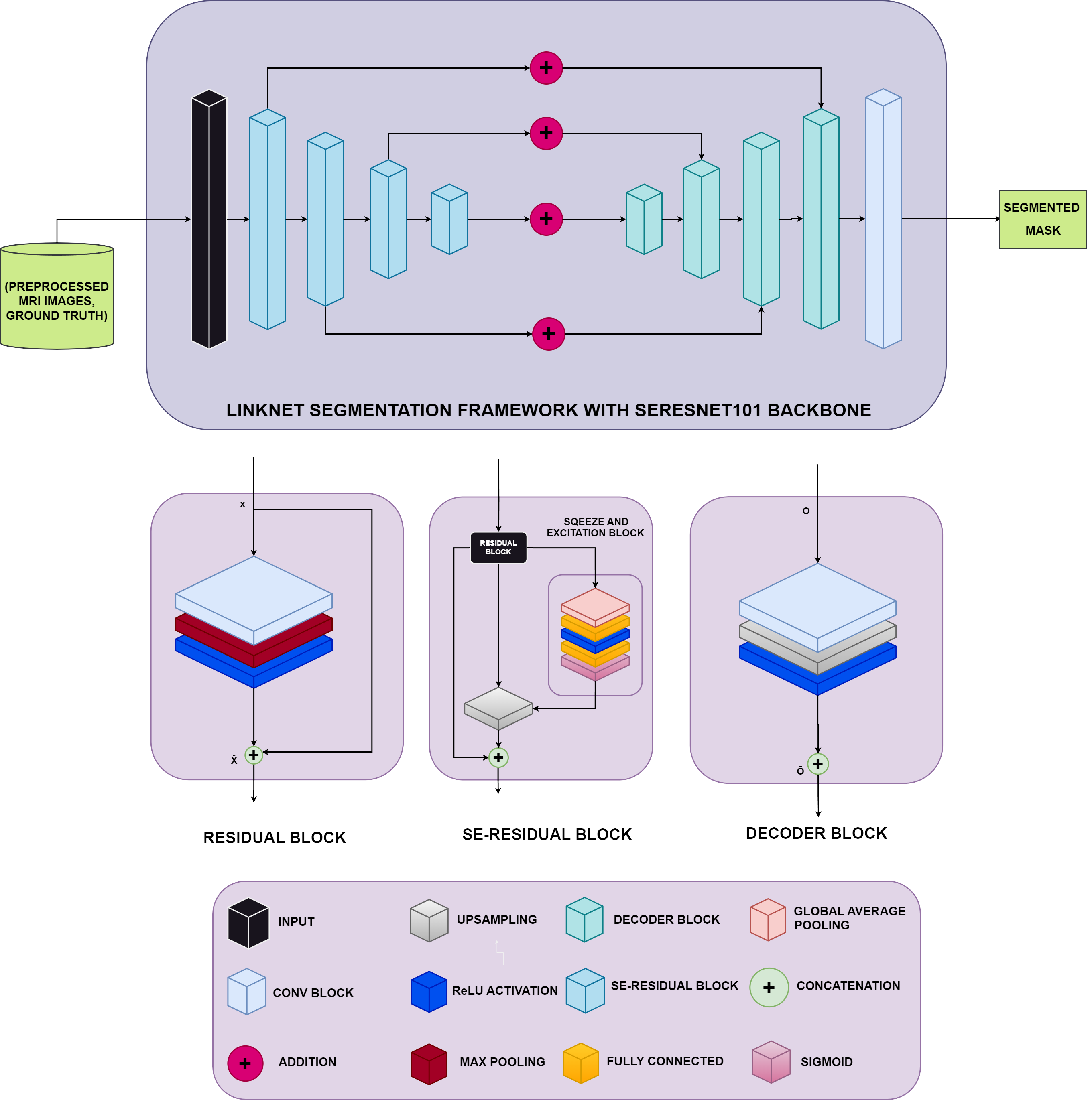}
    \caption{LinkNet Framework with SeResNet101 CNN Backbone Architecture for Brain Tumor Segmentation}
    \label{fig:SegmentationArchitecture}
\end{figure}

\subsection{\textbf{SE-ResNet152 Enhanced Adaptive Boosting Framework for Tumor Classification}}
We propose a 3D brain tumor classification framework built upon a SeResNet152 CNN backbone, integrated with an Adaptive Boosting classifier to accurately differentiate tumor grades (LGG vs. HGG) from segmented MRI volumes. The classification architecture (Figure~7) consists of three core stages: hierarchical feature extraction via SE-Residual blocks, dimensionality reduction via feature flattening, and robust tumor-grade classification leveraging adaptive boosting.

% algo 4

\begin{algorithm}[!h]
\caption{Training SE-ResNet152 with Adaptive Boosting for Brain Tumor Classification}
\label{algo:seresnet_adaboost_classification}
\KwIn{Segmented tumor MRI volume $X\in\mathbb{R}^{H\times W\times D}$, tumor grade labels $Y$, epochs $E$, initial learning rate $\mu$, batch size $\beta$, AdaBoost iterations $\tau$}
\KwOut{Trained SE-ResNet152 and AdaBoost parameters ($\theta, \{\alpha_t,h_t\}_{t=1}^{\tau}$)}

Initialize SE-ResNet152 parameters $\theta$\;

\For{$epoch = 1,\dots,E$}{
    \For{each batch $(X_b,Y_b)$ of size $\beta$}{
        \tcp{Forward pass through SE-Residual Blocks}
        $X^{(0)}\gets X_b$\;
        \For{$l=1,\dots,6$ SE-Residual blocks}{
            $X^{(l)}\gets X^{(l-1)}+SE\left(F_{\text{res}}(X^{(l-1)};W_l)\right)$\;
        }

        \tcp{Flatten extracted features}
        $Z_b\gets\text{Flatten}(X^{(6)})$\;

        \tcp{Compute classification loss (cross-entropy)}
        $L(\theta)\gets -\frac{1}{\beta}\sum_{i=1}^{\beta}Y_b^{(i)}\log(\hat{Y}_b^{(i)})$\;

        \tcp{Backpropagation and parameter update}
        $\theta\gets\theta-\mu\nabla_{\theta}L(\theta)$\;
    }

    Evaluate validation performance; reduce $\mu$ if performance stagnates\;
}

\tcp{Train AdaBoost Classifier}
Initialize sample weights: $w_i^{(1)}\gets\frac{1}{\eta},\,\forall i=1,\dots,\eta$\;

\For{$t=1,\dots,\tau$}{
    Train weak classifier $h_t(Z)$ on extracted features $Z$\;

    Compute classification error:
    $\epsilon_t=\frac{\sum_{i=1}^{\eta}w_i^{(t)}I(h_t(Z_i)\neq Y_i)}{\sum_{i=1}^{\eta}w_i^{(t)}}$\;

    Update classifier weight:
    $\alpha_t=\frac{1}{2}\ln\left(\frac{1-\epsilon_t}{\epsilon_t}\right)$\;

    Update sample weights:
    $w_i^{(t+1)}=w_i^{(t)}\exp(-\alpha_tY_ih_t(Z_i))$\;

    Normalize weights:
    $w_i^{(t+1)}=\frac{w_i^{(t+1)}}{\sum_{j=1}^{\eta}w_j^{(t+1)}}$\;
}

\tcp{Final strong classifier}
Construct final classifier:
$H(Z)=\text{sign}\left(\sum_{t=1}^{\tau}\alpha_th_t(Z)\right)$\;

\Return $\theta$, $\{\alpha_t,h_t\}_{t=1}^{\tau}$\;
\end{algorithm}

Specifically, the MRI volume $X \in \mathbb{R}^{H\times W\times D \times C}$ initially passes through a series of six SE-Residual blocks. Each SE-Residual block combines residual learning with a Squeeze-and-Excitation (SE) module to recalibrate channel-wise feature importance adaptively. This recalibration is particularly valuable for tumor grading, as it emphasizes histologically-relevant imaging biomarkers while suppressing normal tissue variations. For instance, SE modules can highlight necrotic regions, irregular vasculature patterns, and abnormal cell proliferation zones—key radiological indicators that distinguish high-grade from low-grade gliomas.

Given an input feature map $X^{(l)}$ at block $l$, the output is computed as:
\begin{equation}
Y^{(l)} = X^{(l)} + SE\left(F_{\text{res}}(X^{(l)}, \mathbf{W}^{(l)})\right),
\end{equation}
where $F_{\text{res}}(X^{(l)}, \mathbf{W}^{(l)})$ represents the residual mapping parameterized by learnable weights $\mathbf{W}^{(l)}$, and the $SE(\cdot)$ operation adaptively recalibrates channel-wise responses. This residual learning approach preserves both normal brain tissue characteristics and pathological deviations, mirroring how neuro-oncologists assess tumor grade by comparing abnormal regions against normal brain parenchyma.

The deep architecture with six SE-Residual blocks enables hierarchical extraction of increasingly complex tumor features, analogous to the multi-level histopathological criteria used in the WHO grading system. Lower-level blocks detect basic textural abnormalities (corresponding to cellular density variations), while deeper blocks capture complex patterns like necrosis, microvascular proliferation, and infiltrative growth patterns—all critical determinants in distinguishing LGG from HGG. The SE mechanism dynamically emphasizes these grade-specific imaging biomarkers across different patients, accounting for the substantial heterogeneity observed in glioma presentation.

Following feature extraction, the resultant 3D feature representation from the final SE-Residual block $Y^{(6)}$ is flattened into a feature vector $\mathbf{Z} \in \mathbb{R}^{F}$, where:
\begin{equation}
\mathbf{Z} = \text{Flatten}(Y^{(6)}).
\end{equation}
This flattening operation integrates spatial information across the entire tumor volume, similar to how pathologists examine multiple tumor sections to establish an overall grade rather than basing assessment on isolated regions.

The flattened feature vector $\mathbf{Z}$ is then provided as input to an Adaptive Boosting (AdaBoost) classifier, which constructs a robust classifier by iteratively training and combining multiple weak classifiers. Our choice of AdaBoost parallels the clinical decision-making process where multiple diagnostic criteria (analogous to weak classifiers) are weighted and combined to reach a final diagnosis. This is particularly relevant for challenging cases where individual radiological findings might be ambiguous, but their weighted combination yields a definitive grade classification.

% fig
\begin{figure}[!t]
    \centering
    \includegraphics[width=\textwidth]{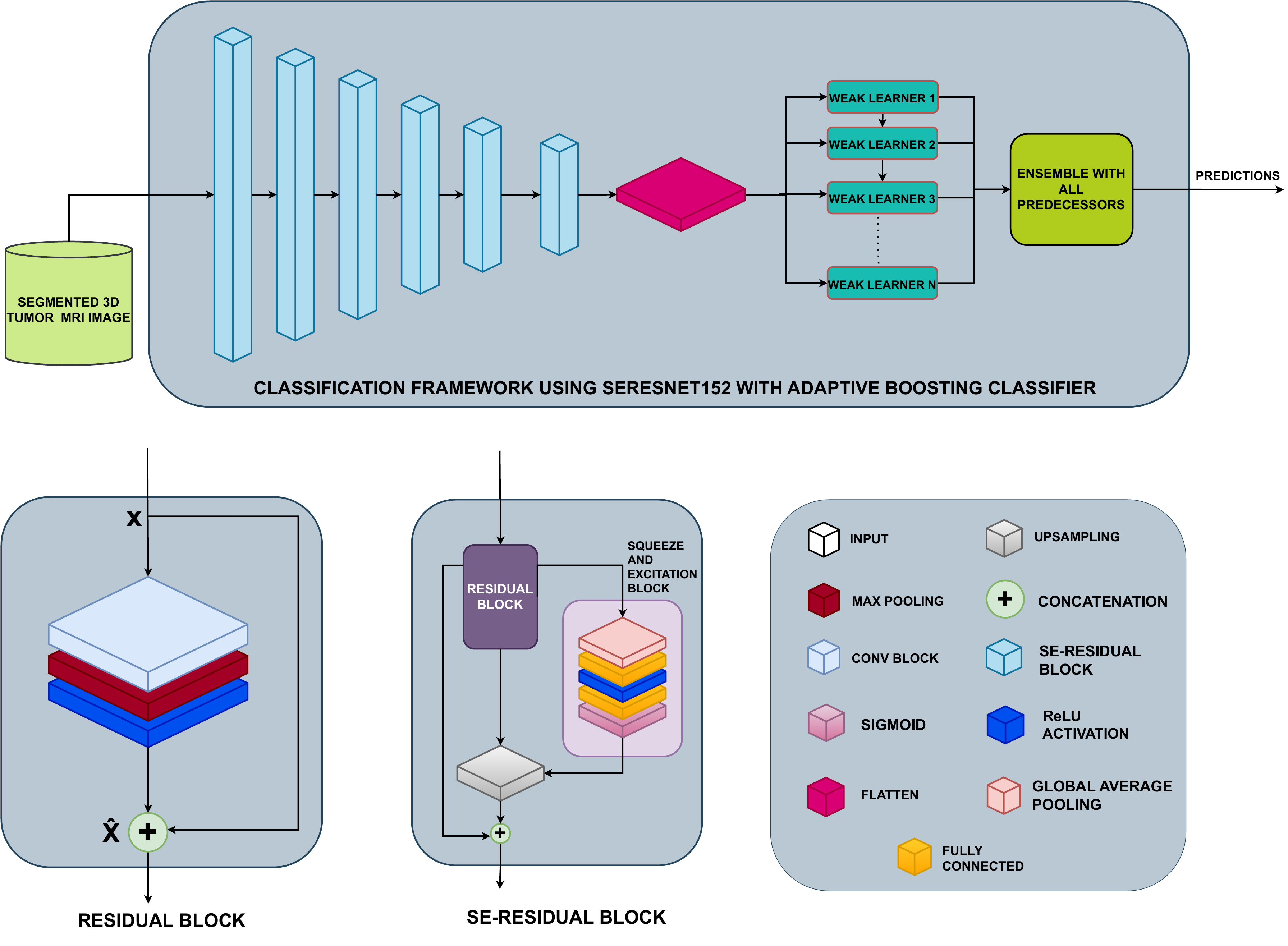}
    \caption{SeResNet152 CNN Backbone Enhanced Adaptive Boosting Framework for Brain Tumor Classification}
    \label{fig:classificationArchitecture}
\end{figure}

Formally, the AdaBoost classifier produces the final tumor classification decision as:
\begin{equation}
H(\mathbf{Z}) = \text{sign}\left(\sum_{t=1}^{T}\alpha_t\,h_t(\mathbf{Z})\right),
\end{equation}
where each weak classifier $h_t(\mathbf{Z})$ contributes to the final classification with weight $\alpha_t$, iteratively adjusted based on misclassification performance. This adaptive weighting gives varying importance to different radiological features for specific tumor subtypes.

The integration of SE-ResNet152-based hierarchical feature extraction with adaptive boosting significantly enhances the classifier's ability to differentiate complex patterns indicative of Low-Grade Gliomas (LGG) and High-Grade Gliomas (HGG). This enhanced differentiation capability is particularly important for identifying molecular and genetic subtypes that are present with subtle radiological differences yet require distinct treatment approaches. 

Thus, our framework effectively models the spectrum of glioma manifestations, from the more indolent growth patterns of LGG to the aggressive, heterogeneous appearance of HGG with their characteristic contrast enhancement, thereby improving tumor-grade classification accuracy in clinical MRI applications. The specifications and parameters of the SeResNet152 CNN backbone-enhanced adaptive boosting classifier are detailed in Table~\ref{tab:seresnet_adaboost_classification_specs}, and the working is presented in Algorithm~\ref{algo:seresnet_adaboost_classification}.

\begin{table}[htbp!]
\centering
\caption{SeResNet152 CNN Backbone Enhanced Adaptive Boosting Classifier Specifications}
\begin{tabular}{lcc}
\hline
\textbf{Parameters} & \textbf{Coefficients} \\ \hline
Epochs              & 100                  \\
Batch Size          & 8                    \\ 
Learning Rate       & 0.001                \\ 
Total Trainable Parameters & 54,560,482        \\ 
Image Shape         & (128, 128, 128)      \\ \hline
\end{tabular}
\label{tab:seresnet_adaboost_classification_specs}
\end{table}

}

\section{Results}{
\subsection{Experimental Setup}
This section presents the performance evaluation of the proposed model compared to existing DL approaches. Our experiments were conducted on a high-performance computing (HPC) system with detailed specifications provided in Table~\ref{tab:sysSpecs}. The system utilizes the Linux 5.15 operating system and features a powerful AMD EPYC 7763 CPU with 128 cores and x86-64 architecture. Additionally, it leverages a dual-socket AMD Radeon Instinct GPU, each socket equipped with 64 cores and runs one thread per core.

\begin{table}[h!]
\centering
\caption{System Specifications for Experimental Setup}
\begin{tabular}{l l}
\hline
\textbf{Component}        & \textbf{Specification}         \\ \hline
CPU                       & AMD EPYC 7763                  \\
Architecture              & x86\_64                        \\
Threads per Core          & 1                            \\
Sockets                   & 2                               \\
Cores per Socket          & 64                              \\
GPU                       & AMD Radeon Instinct            \\
RAM                       & 32GB                            \\
OS                        & Linux 5.15                      \\ \hline
\end{tabular}
\label{tab:sysSpecs}
\end{table}

\subsection{Evaluation Metrics for Segmentation and Localization}

We evaluate the proposed segmentation and localization frameworks using well-established metrics: the Jaccard-Focal loss, Dice coefficient, Intersection over Union (IoU), and Mean IoU.

\textbf{Jaccard-Focal Loss.} To address class imbalance and enhance boundary precision, our segmentation model utilizes a combined loss of focal loss and Jaccard (IoU) loss. Focal loss emphasizes hard-to-classify voxels and is defined as:
\begin{equation}
\mathcal{L}_{\text{focal}}(p,y) = -\alpha(1-p)^\gamma y\log(p)-(1-\alpha)p^\gamma(1-y)\log(1-p),
\end{equation}

where $\alpha$ balances class importance, $\gamma$ controls the focus on hard samples, $p$ is the predicted probability, and $y$ is the ground-truth label. Jaccard loss ($\mathcal{L}_{\text{Jaccard}}$) directly penalizes inaccuracies in overlap between predicted ($V_p$) and ground-truth ($V_g$) volumes, defined as:

\begin{equation}
\mathcal{L}_{\text{Jaccard}} = 1 - \frac{|V_p\cap V_g|}{|V_p\cup V_g|}.
\end{equation}

The total loss used for training combines these two components:
\begin{equation}
\mathcal{L}_{\text{total}} = \mathcal{L}_{\text{focal}} + \mathcal{L}_{\text{Jaccard}}.
\end{equation}

\textbf{Dice Coefficient.} We measure segmentation overlap accuracy using the Dice coefficient, computed as:
\begin{equation}
\text{Dice} = \frac{2\,|V_p\cap V_g|}{|V_p| + |V_g|}.
\end{equation}

\textbf{Intersection over Union (IoU).} Localization and segmentation quality are further assessed by the IoU metric, calculated as the overlap ratio between predicted and ground-truth regions:
\begin{equation}
\text{IoU} = \frac{|V_p\cap V_g|}{|V_p\cup V_g|}.
\end{equation}

\textbf{Mean IoU.} To provide an aggregate performance measure, the Mean IoU computes the average IoU across $N$ samples:
\begin{equation}
\text{Mean IoU} = \frac{1}{N}\sum_{i=1}^{N}\text{IoU}(y_{\text{true}}^{(i)}, y_{\text{pred}}^{(i)}).
\end{equation}

These metrics collectively ensure fair evaluation, focusing explicitly on accuracy, boundary precision, and overall localization and segmentation quality.

\subsection{Localization Performance Evaluation and Discussion}

The localization framework was trained and validated for 100 epochs. These metrics are plotted against their corresponding epochs and are illustrated in Figures \ref{fig:localizationAcc}-\ref{fig:localizationIoU}.

\begin{figure}[h!]
    \centering
    \begin{minipage}[b]{0.48\textwidth}
        \centering
        \includegraphics[width=1\textwidth]{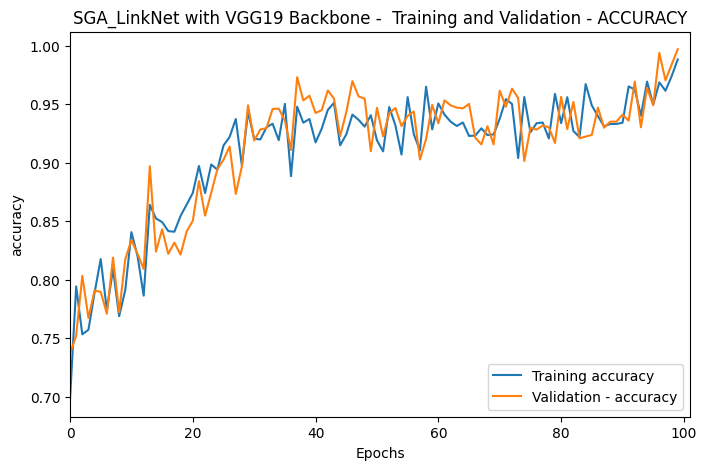} % Adjust width as needed
        \caption{Training and Validation Accuracy Curves of Proposed Localization Framework}
        \label{fig:localizationAcc}
    \end{minipage}
    \hfill
    \begin{minipage}[b]{0.48\textwidth}
        \centering
        \includegraphics[width=1\textwidth]{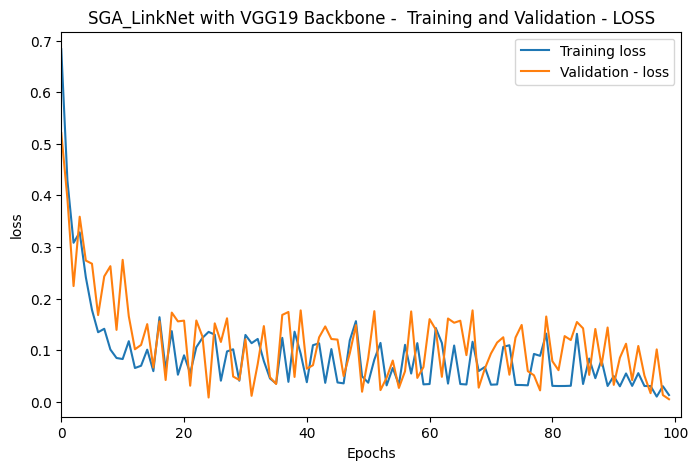}
    \caption{Training and Validation Loss Curves of Proposed Localization Framework}
    \label{fig:localizationLoss}
    \end{minipage}
\end{figure}

\begin{figure}[h!]
    \centering
    \begin{minipage}[b]{0.48\textwidth}
        \centering
        \includegraphics[width=1\textwidth]{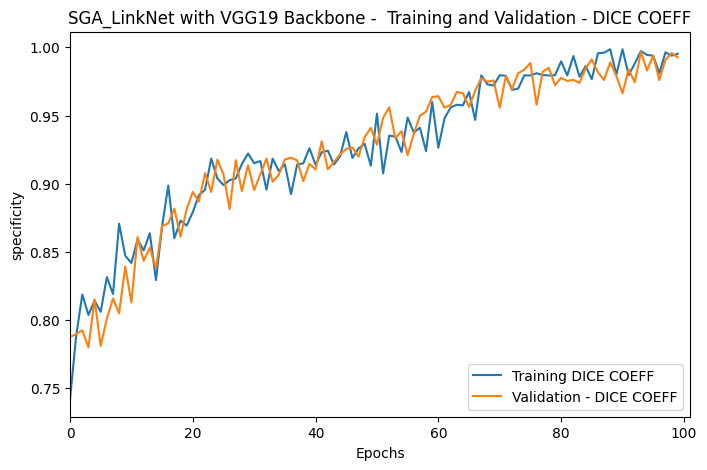} % Adjust width as needed
        \caption{Training and Validation Dice Coefficient Curves of Proposed Localization Framework}
        \label{fig:localizationDice}
    \end{minipage}
    \hfill
    \begin{minipage}[b]{0.48\textwidth}
        \centering
        \includegraphics[width=1\textwidth]{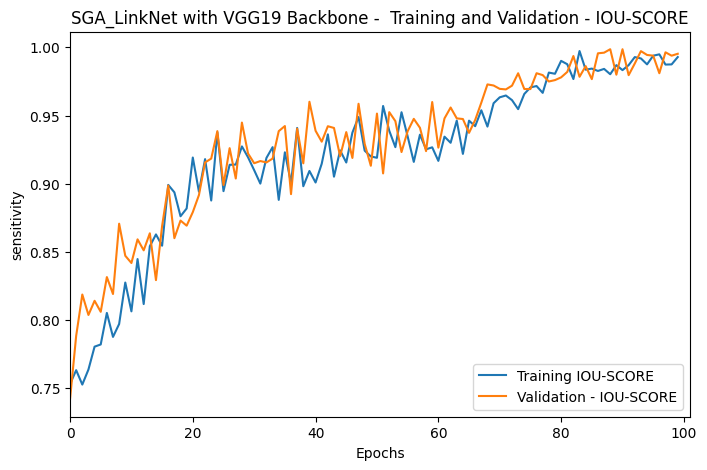} % Adjust width as needed
        \caption{Training and Validation IoU Score Curves of Proposed Localization Framework}
        \label{fig:localizationIoU}
    \end{minipage}
\end{figure}

\begin{figure}[h!]
    \centering
    \includegraphics[width=\textwidth]{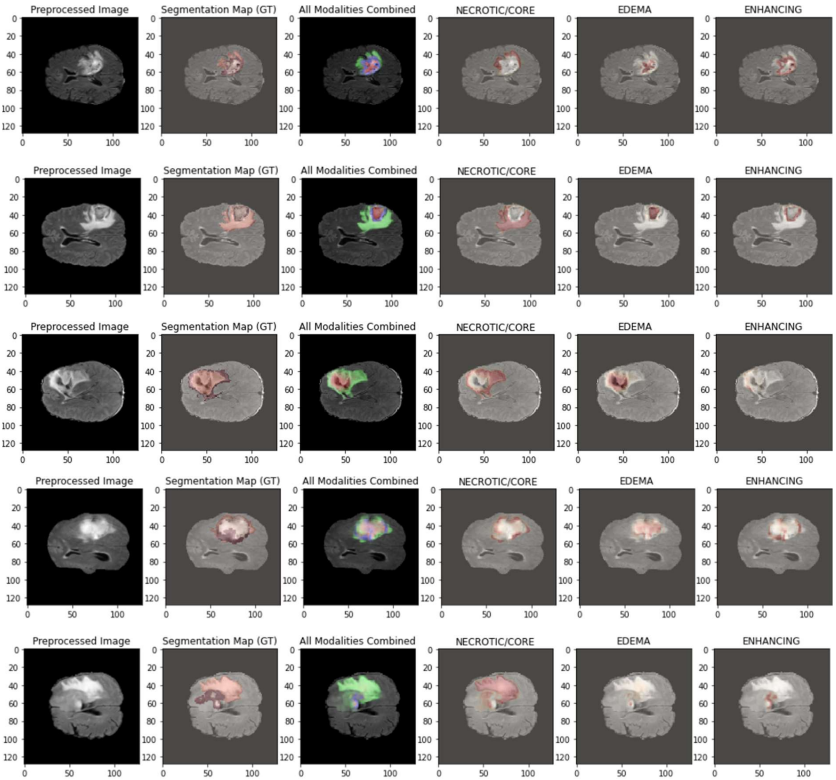}
    \caption{Visual Representation of Tumor Localization from MRI Utilizing Multimodal Cues Using Proposed SGA LinkNet Localization Framework}
    \label{fig:localizationVisualization}
\end{figure}

Figure \ref{fig:localizationVisualization} shows the visual representation of tumor localization from MRI utilizing multimodal cues. The performance of our proposed localization framework was compared with other SOTA models: Shallow CNN~\cite{shallow3DCnn}, ResNet50~\cite{resnet50}, U-Net~\cite{3dUNeT}, DeepLab V3~\cite{deeplabv3}, V-Net~\cite{vnet}, and DeepMedic~\cite{deepmedic}, each of which was also trained from scratch for a fair comparison. Figure \ref{fig:localizationComparison} illustrates this comparison.

\begin{figure}[h!]
    \centering
    \includegraphics[width=\textwidth]{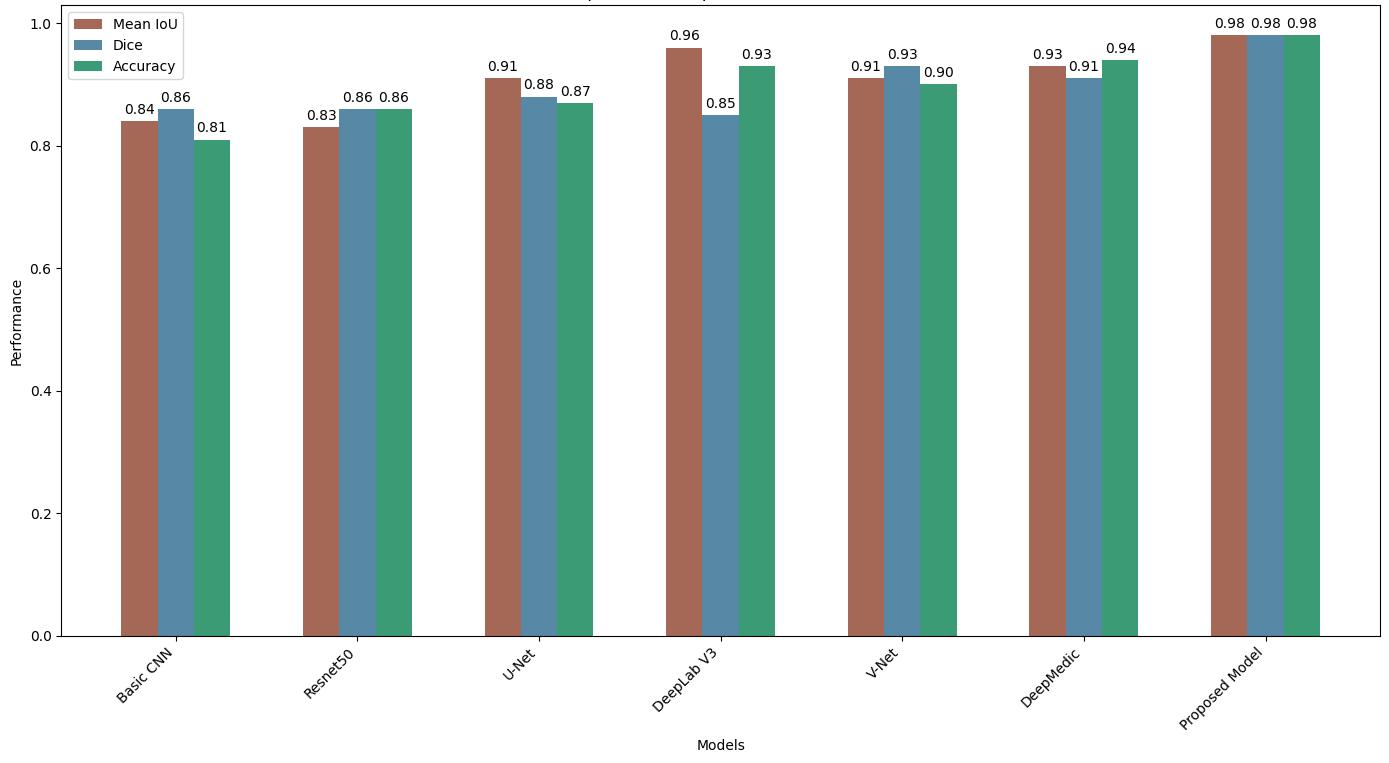}
    \caption{Performance Comparison of SOTA Models for Brain Tumor Localization}
    \label{fig:localizationComparison}
\end{figure}

As seen in Figure \ref{fig:localizationComparison}, our proposed localization framework significantly outperforms the baselines in both the primary and the extended criterion. Basic (shallow) CNN models
suffer from segmentation tasks primarily because they do not accommodate sufficient spatial and semantic information and, therefore, yield suboptimal performance. ResNet50 fails to capture detail from MRI scans that harms the ability of the system to extract meaningful tumor features and yields lower accuracy.

U-Net, with its advanced architecture, suffers from overfitting of its structure and an overwhelming number of parameters, hindering effective generalization across samples. DeepLabV3, while effective at capturing global context, lacks precision in tasks requiring fine-grained localization details, resulting in less accurate boundary delineation and difficulties in delineating small or complex tumor areas. V-Net suffers from its deep architecture due to vanishing gradients, which hampers the learning process in initial layers and impacts its ability to recognize fine details regarding the tumor. DeepMedic's architecture has a limited field of view, causing inconsistencies in localizations, especially at boundaries of structures or complex anatomical areas where a larger context is required for precise localization.

Our proposed SGA LinkNet localization framework outperforms others due to its dual-attended approach in an encoder-decoder framework. The spatial attention helps enhance the zooming of the model into regions of interest for more accurate tumor localizations by refining feature representations according to spatial contexts. This enables it to highlight specific areas within the MRI scans without suppressing less informative details. In addition, the graph attention captures complicated spatial relationships within the neuronal connections among brain regions and leverages them to achieve better localization accuracy of tumors and delineation of boundaries.

\subsection{Segmentation Performance Evaluation and Discussion}

\begin{figure}[h!]
    \centering
    \begin{minipage}[b]{0.48\textwidth}
        \centering
        \includegraphics[width=1\textwidth]{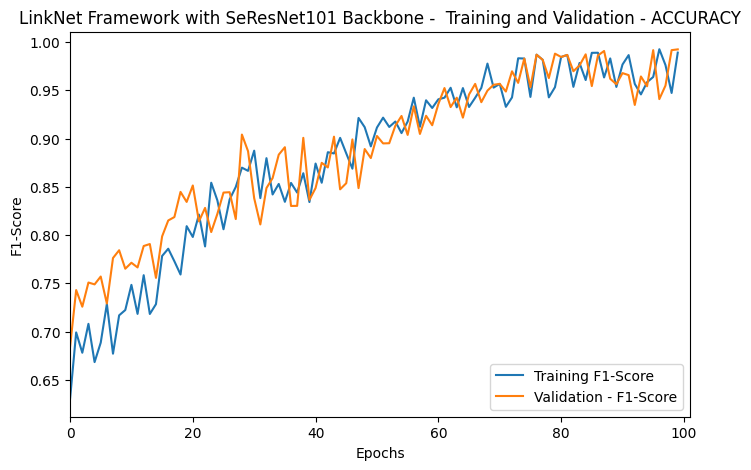} % Adjust width as needed
        \caption{Training and Validation Accuracy Curves of Proposed Segmentation Framework}
        \label{fig:segAcc}
    \end{minipage}
    \hfill
    \begin{minipage}[b]{0.48\textwidth}
        \centering
        \includegraphics[width=1\textwidth]{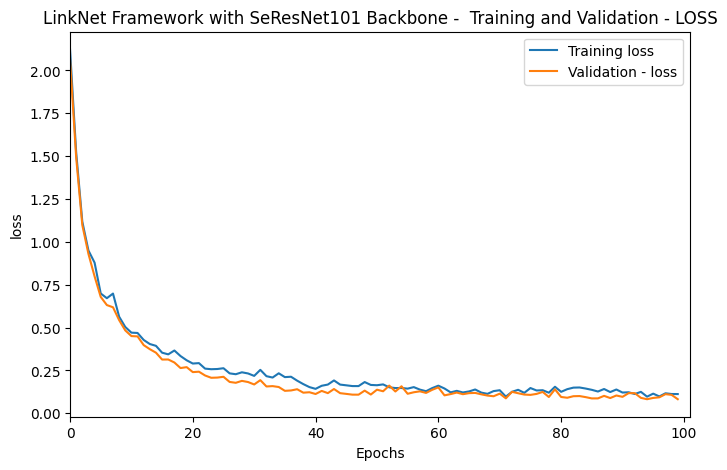} % Adjust width as needed
        \caption{Training and Validation Loss Curves of Proposed Segmentation Framework}
        \label{fig:segLoss}
    \end{minipage}
\end{figure}

\begin{figure}[h!]
    \centering
    \begin{minipage}[b]{0.48\textwidth}
        \centering
        \includegraphics[width=1\textwidth]{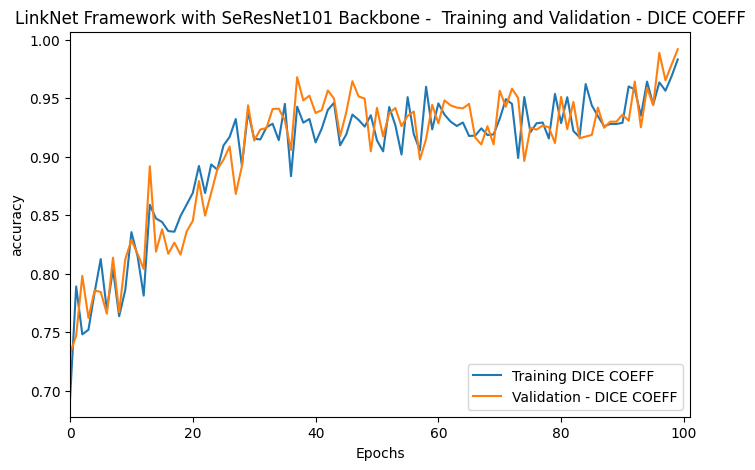} % Adjust width as needed
        \caption{Training and Validation Dice Coefficient Curves of Proposed Segmentation Framework}
        \label{fig:SegDice}
    \end{minipage}
    \hfill
    \begin{minipage}[b]{0.48\textwidth}
        \centering
        \includegraphics[width=1\textwidth]{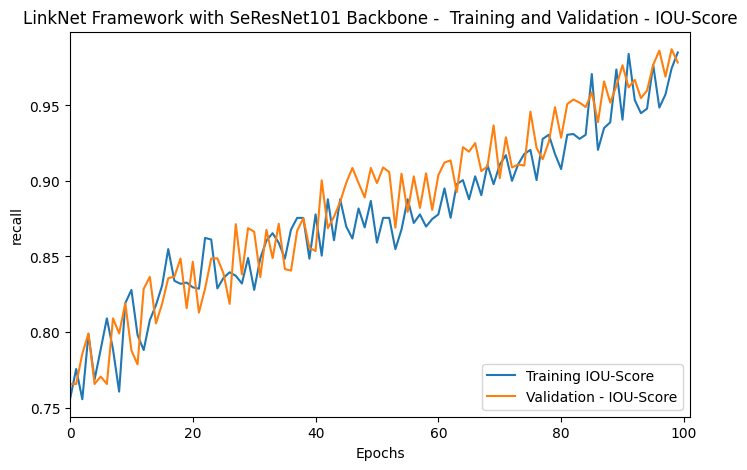} % Adjust width as needed
        \caption{Training and Validation IoU Score Curves of Proposed Segmentation Framework}
        \label{fig:SegIoU}
    \end{minipage}
\end{figure}

\begin{figure}[h!]
    \centering
    \includegraphics[width=0.65\textwidth]{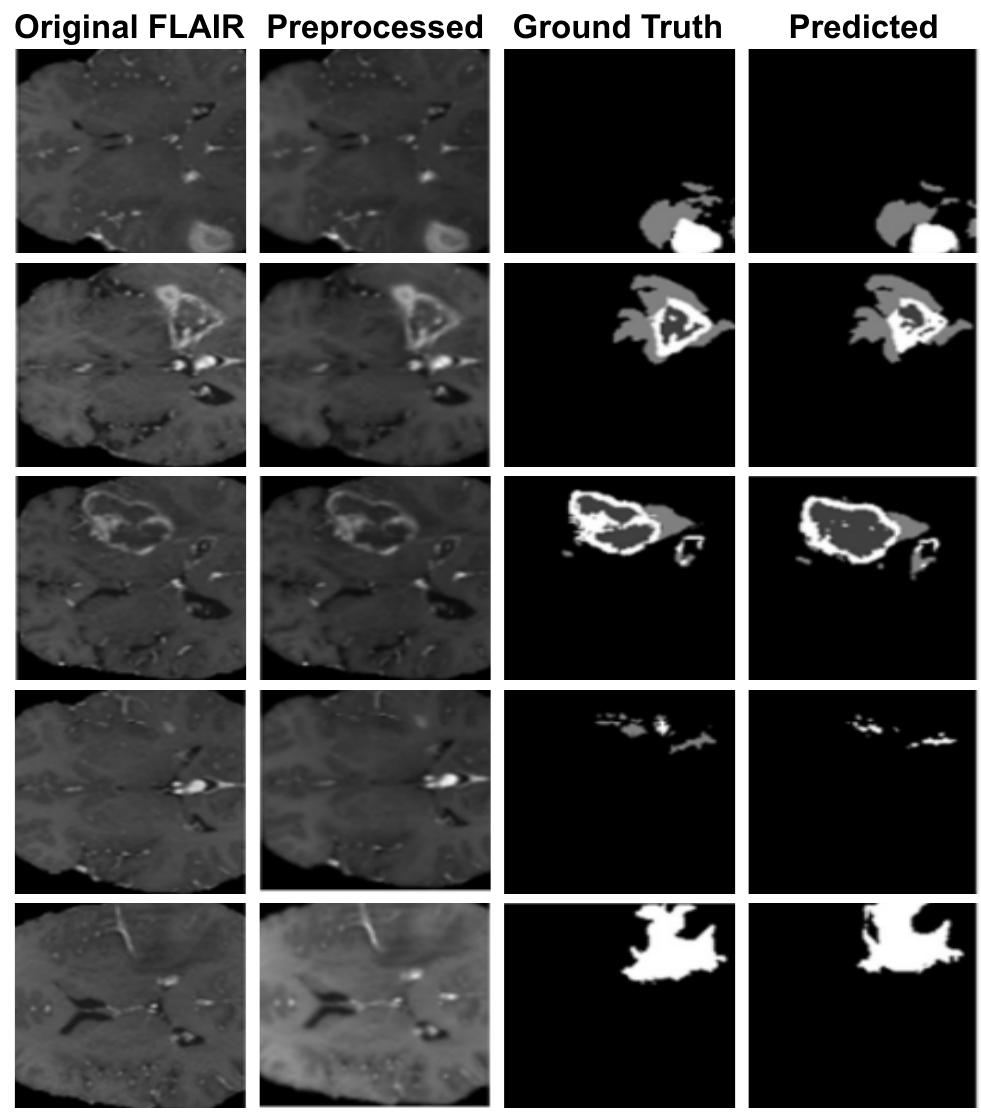}
    \caption{Tumor Segmentation Results Obtained Using Proposed LinkNet Framework with SeResNet101 CNN Backbone.}
    \label{fig:segmentationVisualization}
\end{figure}

Training and validation curves for accuracy and total loss obtained while training the proposed segmentation framework are shown in Figures \ref{fig:segAcc} and \ref{fig:segLoss}. Figure \ref{fig:segmentationVisualization} shows the parallel comparison of the original FLAIR image against the preprocessed, ground truth, and predicted segmentation mask.

The performance of our proposed segmentation framework was benchmarked against the SOTA models: U-Net, 3D U-Net, V-Net, Attention U-Net, nnUNet, DeepMedic, ResUNet, Swin U-Net, DenseVNet, and UResNet, which were also trained with our model to provide a fair and thorough evaluation. The performance of these models is visually compared in Figure \ref{fig:segmentationComparison}.

As illustrated in Figure~\ref{fig:segmentationComparison}, our proposed LinkNet with SE-ResNet101 backbone significantly outperforms existing SOTA segmentation methods. U-Net and 3D U-Net~\cite{3dUNeT} fail to capture detailed tumor boundaries due to limited architectural depth, resulting in lower Dice and IoU scores. VNet~\cite{vnet} and Attention UNet~\cite{3dattunet} suffer from overfitting and high computational demands. Similarly, automatic configuration-based models like DeepMedic~\cite{deepmedic} exhibit inconsistent performance due to suboptimal parameter optimization. Advanced architectures such as ResUNet~\cite{resunet} and Swin U-Net~\cite{swinunet} struggle with precision in overlapping tissue regions, whereas DenseVNet~\cite{DenseVNet} and UResNet~\cite{uresnet}, despite strong feature extraction capabilities, experience slow convergence and reduced accuracy for smaller tumors.

Our proposed LinkNet framework directly addresses these limitations by employing additive skip connections within its encoder-decoder architecture to preserve critical spatial information, enhancing boundary delineation. Additionally, the SE-ResNet101 backbone integrates Squeeze-and-Excitation (SE) modules, dynamically recalibrating channel-wise features, thereby significantly improving the model's sensitivity to important tumor features. This combined approach leads to improved segmentation quality, particularly in capturing fine tumor structures and precise boundaries.

\begin{figure}[h!]
    \centering
    \includegraphics[width=\textwidth]{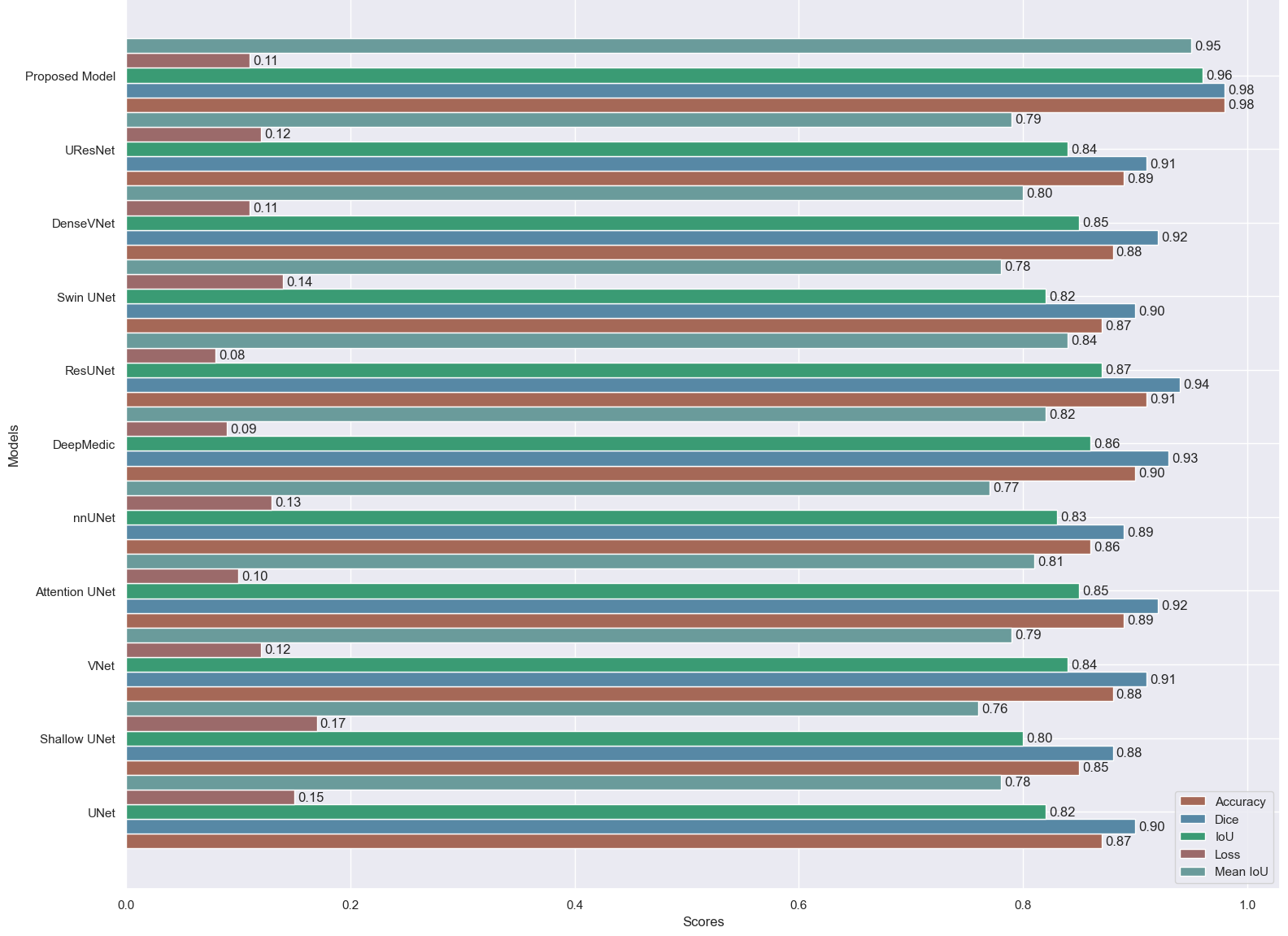}
    \caption{Performance Comparison of SOTA Models for Brain Tumor Segmentation}
    \label{fig:segmentationComparison}
\end{figure}

\subsection{Classification Evaluation Metrics}

We evaluate the proposed tumor classification framework using standard metrics, including Accuracy, Precision, Recall, and F1-score, providing a comprehensive performance assessment in distinguishing Low-Grade Gliomas (LGG) from High-Grade Gliomas (HGG).

\textbf{Accuracy} measures the fraction of correctly classified instances and is computed as:
\begin{equation}
\text{Accuracy} = \frac{TP + TN}{TP + TN + FP + FN},
\end{equation}
where $TP$, $TN$, $FP$, and $FN$ denote true positives (HGG correctly classified), true negatives (LGG correctly classified), false positives (LGG misclassified as HGG), and false negatives (HGG misclassified as LGG), respectively.

\textbf{Precision} quantifies the fraction of correctly classified HGG cases out of all predicted HGG cases, emphasizing minimal false positives:
\begin{equation}
\text{Precision} = \frac{TP}{TP + FP}.
\end{equation}

\textbf{Recall} (Sensitivity) evaluates the model’s capability to correctly detect actual HGG cases, critical in avoiding under-treatment:
\begin{equation}
\text{Recall} = \frac{TP}{TP + FN}.
\end{equation}

\textbf{F1-score} represents the harmonic mean of Precision and Recall, balancing false positives and negatives:
\begin{equation}
\text{F1-score} = 2 \cdot \frac{\text{Precision} \cdot \text{Recall}}{\text{Precision} + \text{Recall}}.
\end{equation}

\subsection{Classification Performance Evaluation and Discussion}

Our proposed classification framework was trained for 100 epochs, and the training-validation curves for all metrics are illustrated in Figures~\ref{fig:clsAcc}, \ref{fig:clsF1}, \ref{fig:clsPrec}, \ref{fig:clsRec}, and \ref{fig:clsnLoss}.

\begin{figure}[h!]
    \centering
    \begin{minipage}[b]{0.48\textwidth}
        \centering
        \includegraphics[width=1\textwidth]{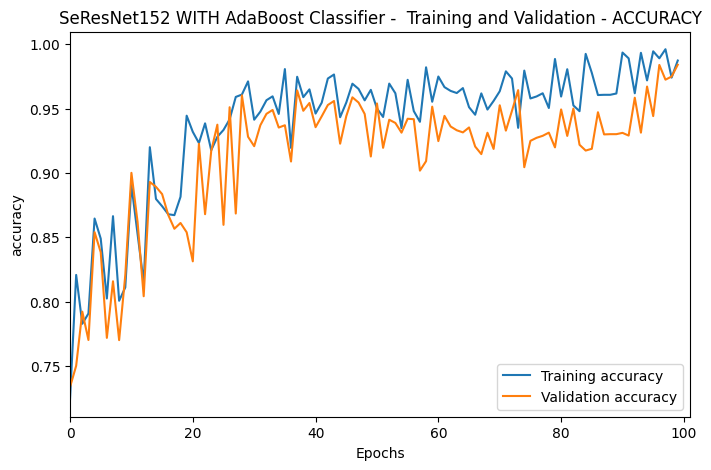} % Adjust width as needed
        \caption{Epoch-Wise Accuracy Scores during Training and Validation Phases of SeResNet152 Neural Network Backbone Enhanced Adaptive Boosting Framework}
        \label{fig:clsAcc}
    \end{minipage}
    \hfill
    \begin{minipage}[b]{0.48\textwidth}
        \centering
        \includegraphics[width=1\textwidth]{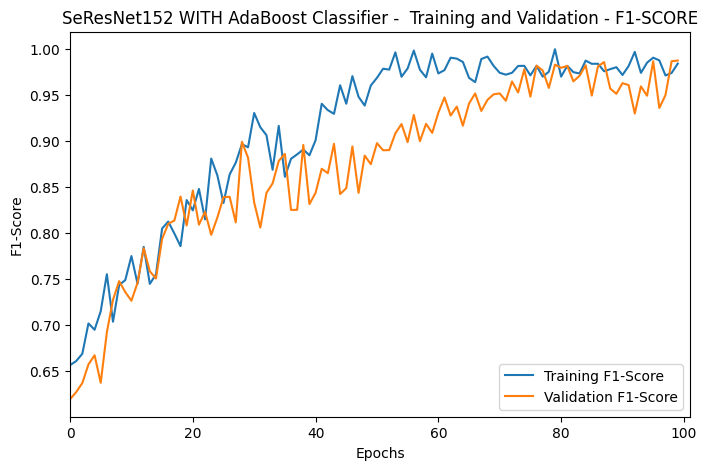} % Adjust width as needed
        \caption{Epoch-Wise F1 Scores during Training and Validation Phases of SeResNet152 Neural Network Backbone Enhanced Adaptive Boosting Framework}
        \label{fig:clsF1}
    \end{minipage}
\end{figure}

\begin{figure}[h!]
    \centering
    \begin{minipage}[b]{0.48\textwidth}
        \centering
        \includegraphics[width=1\textwidth]{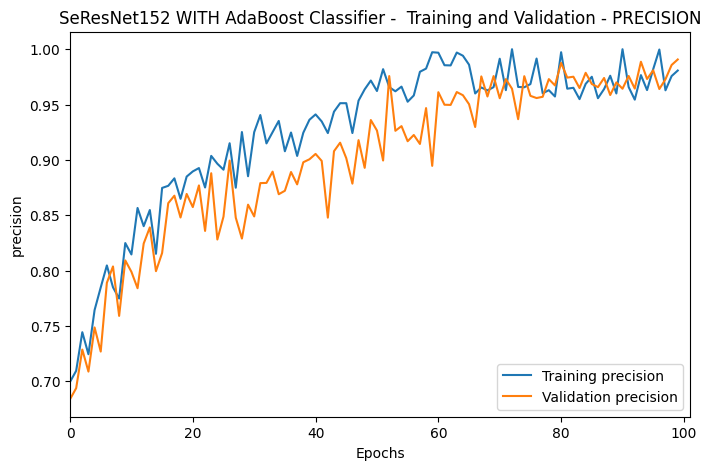} % Adjust width as needed
        \caption{Epoch-Wise Precision Scores during Training and Validation Phases of SeResNet152 Neural Network Backbone Enhanced Adaptive Boosting Framework}
        \label{fig:clsPrec}
    \end{minipage}
    \hfill
    \begin{minipage}[b]{0.48\textwidth}
        \centering
        \includegraphics[width=1\textwidth]{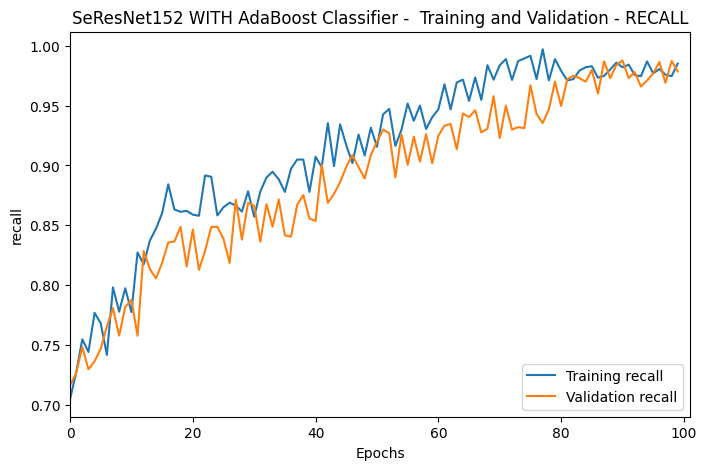} % Adjust width as needed
        \caption{Epoch-Wise Recall Scores during Training and Validation Phases of SeResNet152 Neural Network Backbone Enhanced Adaptive Boosting Framework}
        \label{fig:clsRec}
    \end{minipage}
\end{figure}

\begin{figure}[h!]
    \centering
    \begin{minipage}[b]{0.48\textwidth}
        \centering
        \includegraphics[width=1\textwidth]{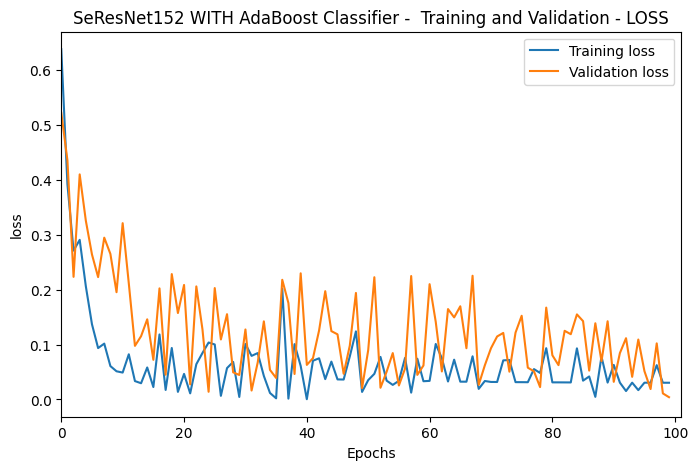}
    \caption{Epoch-Wise Loss Curves during Training and Validation Phases of SeResNet152 Neural Network Backbone Enhanced Adaptive Boosting Framework}
    \label{fig:clsnLoss}
    \end{minipage}
    \hfill
    \begin{minipage}[b]{0.48\textwidth}
        \centering
        \includegraphics[width=1\textwidth]{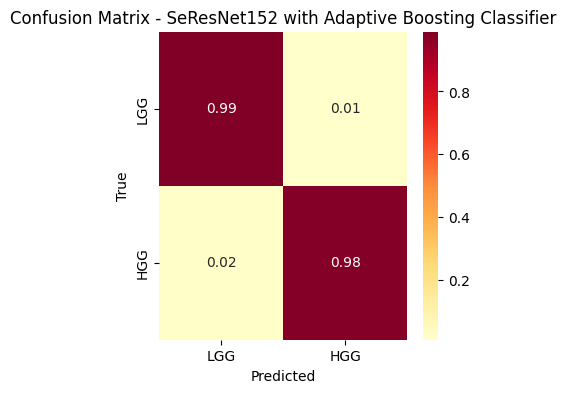}
    \caption{Confusion Matrix Obtained from Proposed SeResNet152 CNN Backbone Enhanced Adaptive Boosting Framework}
    \label{fig:clsCM}
    \end{minipage}
\end{figure}

Figure~\ref{fig:clsCM} presents the confusion matrix for our classification framework, where rows represent the actual tumor classes (LGG and HGG), and columns represent the predicted classes. Diagonal elements indicate correct classifications, while off-diagonal elements represent misclassifications. The matrix proves that our model accurately classifies 99\% of LGG cases and 98\% of HGG cases, with minimal misclassification rates of only 1\% (LGG to HGG) and 2\% (HGG to LGG), respectively.

Our proposed classification framework's performance was benchmarked against other SOTA models, including ResNet50~\cite{resnet50}, VGG16~\cite{27}, DenseNet121~\cite{densenet}, InceptionV3~\cite{inception}, EfficientNet~\cite{effnet}, Xception~\cite{xception}, InceptionResNetV2~\cite{inceptionResNet}, and MobileNetV3~\cite{mobilenetv3}, all of which were trained alongside our model to ensure a fair and comprehensive evaluation.

\begin{table}[h!]
\centering
\renewcommand{\arraystretch}{1.3} % Row height
\setlength{\tabcolsep}{4pt} % Column spacing

\resizebox{\textwidth}{!}{
\begin{tabular}{lcccccccc}
\hline
\multirow{2}{*}{\textbf{Model}} & \multicolumn{4}{c}{\textbf{Training Performance (\%)}} & \multicolumn{4}{c}{\textbf{Validation Performance (\%)}} \\ \cline{2-9} 
                                & \textbf{Accuracy} & \textbf{Precision} & \textbf{Recall} & \textbf{F1 Score} & \textbf{Accuracy} & \textbf{Precision} & \textbf{Recall} & \textbf{F1 Score} \\ \hline
ResNet50~\cite{resnet50}                        & 97.69             & 95.22              & 97.17           & 96.67             & 94.12             & 96.27              & 96.13           & 96.12             \\ 
VGG16~\cite{27}                             & 97.18             & 95.14              & 96.78           & 96.65             & 95.16             & 95.21              & 96.13           & 95.65             \\ 
DenseNet121~\cite{densenet}                       & 97.39             & 96.41              & 97.27           & 96.60             & 96.14             & 96.12              & 97.12           & 96.43             \\ 
InceptionV3~\cite{inception}                       & 97.23             & 96.28              & 96.81           & 96.54             & 93.23             & 94.23              & 93.16           & 93.14             \\ 
EfficientNet~\cite{effnet}                      & 97.06             & 96.63              & 97.06           & 96.45             & 97.12             & 96.82              & 96.75           & 96.75             \\ 
Xception~\cite{xception}                          & 97.74             & 97.65              & 97.74           & 97.61             & 91.35             & 91.01              & 91.18           & 91.05             \\ 
InceptionResNetV2~\cite{inceptionResNet}                 & 97.52             & 97.42              & 97.53           & 97.45             & 89.16             & 89.81              & 89.91           & 89.85             \\ 
MobileNetV3~\cite{mobilenetv3}                       & 97.41             & 97.31              & 97.42           & 97.35             & 85.27             & 84.82              & 84.91           & 84.85             \\ 
\textbf{Ours}         & \textbf{98.75}    & \textbf{96.91}     & \textbf{97.21}  & \textbf{96.87}    & \textbf{98.53}    & \textbf{97.22}     & \textbf{96.92}  & \textbf{97.34}    \\ \hline
\end{tabular}
}
\caption{Performance comparison of models during training and validation.}
\label{tab:clsComp}
\end{table}

The classification results in Table~\ref{tab:clsComp} demonstrate that our proposed SE-ResNet152-based AdaBoost framework consistently outperforms baseline CNN models for MRI-based brain tumor classification. While ResNet50 struggles to identify subtle tumor invasion patterns due to limited sensitivity to complex spatial features, VGG16’s shallow structure fails to capture critical tumor boundary details, reducing its accuracy. DenseNet121, despite effective gradient flow, does not adequately model complex local tumor features. Similarly, InceptionV3’s global context impairs its ability to precisely capture subtle morphological features necessary for accurate classification.

EfficientNet tends toward overfitting and has limited capability in detecting minor morphological variations and edema regions. Xception insufficiently captures critical spatial relationships required for distinguishing specific tumor indicators like contrast-enhancing regions and necrotic cores. InceptionResNetV2’s highly detailed design paradoxically hinders its sensitivity to subtle features like tumor necrosis and non-enhancing regions. MobileNetV3, prioritizing computational efficiency, lacks sufficient depth to model intricate morphological changes, impairing its early-stage tumor recognition capability.

In contrast, our proposed framework leverages six SE-Residual blocks within an SE-ResNet152 backbone, effectively recalibrating channel-wise features to emphasize tumor boundaries, textures, and subtle morphological variations. The integration of an AdaBoost classifier enhances classification performnce by iteratively correcting misclassifications through a combination of weak learners. Our method achieves a relatively higher performance by effectively capturing critical and subtle tumor features, clearly distinguishing LGG from HGG cases.

}

\section{Conclusion and Future Work}{
    This work presents a multi-stage deep learning framework for automated glioma analysis, integrating preprocessing, localization, segmentation, and classification. We introduce a hierarchical preprocessing pipeline that enhances multimodal MRI volumes through Multiresolution Harmonic Fusion, Adaptive Focused Region Clipping, Luminance-Guided Contrast Enhancement, Dynamic Contextual Smoothing, and Statistical Feature Normalization, ensuring effective feature extraction and noise suppression.

For precise tumor localization, we propose an enhanced LinkNet architecture with a VGG19-inspired encoder, leveraging spatial and graph attention mechanisms to improve multimodal feature integration. Our segmentation framework, built on SeResNet101 within a modified LinkNet backbone, achieves a high IoU of 96\%, enabling accurate delineation of glioma boundaries. Additionally, we develop a classification model that combines SeResNet152 feature extraction with Adaptive Boosting, achieving 98.53\% accuracy in glioma grading. Experimental evaluations demonstrate that our framework consistently outperforms existing methods across all stages, from preprocessing to classification.

Future research can enhance this framework by integrating state-of-the-art deep learning architectures to further refine feature extraction and prediction accuracy. Incorporating patient-specific data, such as genetic profiles and clinical history, could enable personalized diagnosis and treatment planning, advancing towards precision medicine. Additionally, improving real-time processing capabilities for intraoperative and emergency settings remains a key challenge. Deploying this framework on optimized hardware, such as edge AI accelerators, could enable real-time inference, facilitating rapid clinical decision-making. These directions collectively aim to bridge the gap between AI research and real-world clinical applications, moving towards robust, interpretable, and clinically deployable glioma analysis systems.

}

% \section*{Statements and Declarations:}{
% The authors confirm that they have read, understood, and adhered to the ethical responsibilities of authors.
% }

% \section*{Conflict of Interests:}{
% The authors declare that there are no conflicts of interest associated with this work.
% }

% \section*{Funding Information:}{
% This research did not receive any financial support or funding.
% }

% \section*{Data Availability and Access:}
% The results presented in this study are based on the BraTS 2020 dataset, which is publicly available at \color{green}{\textit{\href{https://www.med.upenn.edu/cbica/brats2020/data.html}{https://www.med.upenn.edu/cbica/brats2020/data.html}}}

% \color{black}

%%%%%%%%%%%%%%%%%%%%%%%%%%%%%%%%%%%%%%%%%%%%%%%%%%%%%%%%%%%%%%%%%%%%%%%%%%
%% This file was autogenerated by PaperShell v2.6.1 on 2023-02-03 13:51:10
%% https://github.com/sylvainhalle/PaperShell
%% DO NOT EDIT!
%%%%%%%%%%%%%%%%%%%%%%%%%%%%%%%%%%%%%%%%%%%%%%%%%%%%%%%%%%%%%%%%%%%%%%%%%%
\bibliographystyle{elsarticle-num}

\section*{References}

\bibliography{main}

\end{document}